\documentclass[12pt, reqno]{amsart}
\usepackage{amsmath,amssymb,mathtools,subfigure}
\usepackage{graphicx,color}
\usepackage{tikz,ifthen,cases}
\usetikzlibrary{patterns}
\usepackage{cite}
\usepackage[margin=2.5cm]{geometry}
\usepackage{epsfig}
\usepackage{amsfonts}

\def\NL{{ M }}  
\def\Nv{{ N_{\rm v} }} 
\def\Nc{{ N_{\rm c} }} 
\def\L{{ L }} 
\def\Jump{{ J }} 
\def\Eb{{ E_{\rm b} }} 
\def\Eo{{ E_0 }} 
\def\Ld {{ a }} 
\def\tt {{ \tau }} 
\def\t {{ t }} 

\def\E{{ \mathbb{E} }} 
\def\rhoc{{\bar{\rho}_{\rm crit}}}
\def\Fbar{{\langle F \rangle}}
\def\vbar{{\langle v \rangle}}

\begin{document}

\title[Traffic Flow Models]{On a class of new nonlocal traffic flow models with look-ahead rules}

\author[Y. Sun]{Yi Sun}
\address[Yi Sun]{\newline Department of Mathematics, \ 
 University of South Carolina, 1523 Greene St., Columbia, SC 29208, USA}
\email{yisun@math.sc.edu}
\author[C. Tan]{Changhui Tan}
\address[Changhui Tan]{\newline Department of Mathematics, \ 
 University of South Carolina, 1523 Greene St., Columbia, SC 29208, USA}
\email{tan@math.sc.edu}



\date{}
\maketitle

\begin{abstract}
This paper presents a new class of one-dimensional (1D) traffic models with look-ahead rules that take into account of two effects: nonlocal slow-down effect and right-skewed non-concave  asymmetry in the fundamental diagram. The proposed 1D cellular automata models with the Arrhenius type look-ahead interactions implement stochastic rules for cars' movement following the configuration of the traffic ahead of each car. In particular, we take two different look-ahead rules: one is based on the distance from the car under consideration to the car in front of it; the other one depends on the car density ahead. Both rules feature a novel idea of multiple moves, which plays a key role in recovering the non-concave flux in the
macroscopic dynamics. Through a semi-discrete mesoscopic stochastic process, we derive the coarse-grained macroscopic dynamics of the CA model. We also design a numerical scheme to simulate the proposed CA models with an efficient list-based kinetic Monte Carlo (KMC) algorithm. Our results show that the fluxes of the KMC simulations agree with the coarse-grained macroscopic averaged fluxes for the different look-ahead rules under various parameter settings.

\end{abstract}


\section{Introduction}
The study on traffic flows has received considerable effort in the past few decades.
Many mathematical frameworks and models have been proposed and analyzed in the literature \cite{NER00, CSS00, Hel01, Nag02, Scha02, Ker04, MaD05, KuG11, BeD11, SCN11, TrK13}. These models can be categorized by different scales.

The \emph{microscopic models} focus on modeling the behaviors of individual cars. Two types of microscopic models are widely used in traffic flows.

(i). \emph{The agent-based models}. The location of each car is traced in time, by an interacting ODE system. A variety of interaction rules are proposed, followed by analysis and simulations. In the classical \emph{follow-the-leader} model \cite{GHR61}
\[\dot{v}_i=\kappa_i(v_{i+1}-v_i),\]
the driver in the $i$-th car accelerates/decelerates according to the relative velocity towards the $(i+1)$-th car in front.
In the {\it optimal velocity} (OV) model \cite{BHH95}
\[\dot{v}_i=\kappa_i(V_i-v_i),\]
where $V_i$ is the optimal velocity of the $i$-th car, which depends on the distance toward the car in front. The OV model has several extensions. In \cite{LWS99,Nag99}, the optimal velocity is modeled by multiple cars in front. In \cite{NSH02, HNS03, HNS04}, cars behind are also taken into consideration. Ref.\cite{WBH04} discussed and analyzed multiple look-ahead models, which reproduce many real flow features. Besides the common fundamental properties of traffic flow, free flow and congested flow, all these models above can reproduce a third fundamental property of traffic flow, the so-called synchronized flow \cite{KeR97}.

(ii) \emph{The lattice models}. The road is configured as a fixed lattice. Each site has values $1$ (car is present) or $0$ (car is absent). Explicit rules for car movement on the lattice sites are described to represent the traffic flow. The lattice models, also known as \emph{cellular automata} (CA) models \cite{Wol86, Wol94}, have been widely used to represent traffic flows. A vast literature exists addressing various analytical and numerical techniques for models of this type \cite{CrL86, NaS92, BML92, NaP95, BSS98, KSS00, KSS02, LiT05, Nel06}. Compared with the agent-based models, lattice models are simpler to implement and are more amenable to numerical investigation. A major advantage of lattice models is that they  allow for a systematic derivation of the coarse-grained dynamics, which typically requires some simplifying assumptions about the statistical behavior of the microscopic model.

To understand the emergent phenomena for large crowded traffic systems, a variety of \emph{macroscopic models} have been proposed. Instead of working on individual cars, these models describe the evolution of the density distribution of the traffic. One classical model in 1D with a single-lane dynamics is the Lighthill-Whitham-Richards (LWR) model \cite{LiW55},
\begin{equation}\label{eq:LWR}
  \partial_t\rho+\partial_x(\rho u)=0,\quad u=u_{\max}(1-\rho).
\end{equation}
Here, $\rho$ is the density of the traffic and $u$ is the macroscopic velocity, which takes maximum value $u_{\max}$ if $\rho=0$, and becomes 0 if the maximum density $\rho=1$ is reached. It is well-known that the Burgers-type nonlinearity leads to a finite-time wave breakdown  for any smooth initial configuration. This corresponds to the creation of traffic jams.

The LWR model \eqref{eq:LWR} has many extensions. One direction is to consider the \emph{nonlocal slowdown effect}: drivers intend to slow down if heavy traffic is ahead. This would involve a nonlocal interaction with a look-ahead distance $a$,
\begin{equation}\label{eq:SK}
  \partial_t\rho+\partial_x(\rho u)=0,\quad u=u_{\max}(1-\rho)\exp\left[-\int_0^aK(y)\rho(x+y)dy\right],
\end{equation}
where $K$ is a look-ahead kernel. The model was first introduced by Sopasakis and Katsoulakis (SK) in \cite{SKa06}, with $K\equiv1$. Another kernel $K(r)=2(1-\frac{r}{a})$ was discussed in \cite{CKPT} for pedestrian flows, followed by an extensive numerical study.

The wave breakdown phenomenon for the SK model \eqref{eq:SK} and related nonlocal models has been studied in \cite{LL15, Lee18}. Recently, it is shown in \cite{LT19} that the nonlocal slowdown effect can help avoid traffic jams for a family of initial configurations.

Another concern on the LWR model \eqref{eq:LWR} is on its \emph{fundamental diagram}. The flux
\[F=\rho u=u_{\max}\rho(1-\rho)\]
is a concave function of $\rho$ with an even symmetry at $\rho=1/2$, which does not agree with the experimental data \cite{SCN11, KuG11}. A better fit would be a right-skewed non-concave flux, which looks like
\begin{equation}\label{eq:fluxcc}
  F=u_{\max}\rho(1-\rho)^\Jump,\quad \Jump>1.
\end{equation}

To take into account of the two effects, we propose a
new class of macroscopic traffic models with the form
\begin{equation}\label{eq:main}
  \partial_t\rho+\partial_x(\rho u)=0,\quad u=u_{\max}(1-\rho)^\Jump\exp\left[-\int_0^aK(y)\rho(x+y)dy\right].
\end{equation}
For $\Jump=1$, it reduces to the SK model \eqref{eq:SK}. For $\Jump=2$, it has been recently introduced and discussed in \cite{Lee19}. The behavior of the solution is different from the SK model, due to the non-concavity of the flux. Yet, the nonlocal slowdown effect could still help avoid finite-time wave breakdown for some initial configurations.

It has been a very active area to study the connections between the microscopic and macroscopic dynamics. From the microscopic models, one can let the number of cars go to infinity, and derive a mean-field limit. The resulting \emph{mesoscopic models} are characterized by kinetic equations. The macroscopic dynamics can then be obtained through appropriate hydrodynamic limits.

In this paper, we focus on the lattice models. In \cite{SKa06}, a CA model with Arrhenius type look-ahead interactions was proposed (see Sec. \ref{sec:micro} for the description in detail). Through a \emph{semi-discrete mesoscopic stochastic process}, the SK model \eqref{eq:SK} can be formally derived as a coarse-grained description of the CA model. Extensions to multilane and multiclass traffic have also been developed in \cite{DuS07, AlS08}. An improved mesoscopic model has been discussed in \cite{HST14}.  Two different look-ahead rules in both one-dimensional (1D) and two-dimensional (2D) CA models were compared in \cite{SuT14}.

A natural question that arises is whether there is a lattice model that connects our new macroscopic model \eqref{eq:main}.

We propose a CA model with two different types of look-ahead rules. The rules feature a novel idea of \emph{multiple moves}, which plays a key role in recovering the non-concave  flux in the macroscopic dynamics, for any $\Jump\in\mathbb{Z}_+$.
The first rule has a slowdown factor that is based on the distance from the car under consideration to the car in front of it. The corresponding macroscopic dynamics is a local scalar conservation law with non-concave flux \eqref{eq:fluxcc}. The other rule's slowdown factor depends on the car density ahead. With such long-range interaction, we derive the target coarse-grained macroscopic dynamics \eqref{eq:main}.

We also design a numerical scheme for the proposed CA models.
To improve computational efficiency, we use the kinetic Monte Carlo (KMC) algorithm \cite{BKL75} due to its main feature ---``rejection-free''. When the dynamics of the traffic system features a finite number of distinct processes in configurational changes, we develop an efficient list-based KMC algorithm using a fast search that can further improve the efficiency compared to the general KMC method. On the other hand, the Metropolis Monte Carlo (MMC) method \cite{MRR53} is adopted in most of current CA models for vehicular flows and pedestrian flows. But the MMC method is a way of simulating an equilibrium distribution for a model, and trial steps are sometimes rejected because the acceptance probability is small, in particular when a system approaches the equilibrium, or the car density is high. Therefore, we choose the KMC, which is more suitable for simulating the time evolution of the traffic systems with the transition rates that are associated with possible configurational changes in the system. With reasonable values of the model parameters (the characteristic time unit and the interaction strength), the KMC simulations are used to predict the time evolution of 1D traffic flows. Our results show that the rules induce nonlocal slow-down effect and right-skewed non-concave asymmetry in the fundamental diagram.
Moreover, the fluxes of the KMC simulations agree with the coarse-grained macroscopic averaged fluxes for the different look-ahead rules under various parameter settings.

The rest of the paper is organized as follows.
In Sec. \ref{sec:micro}, we introduce the CA models with two look-ahead rules.
In Sec. \ref{sec:macro}, we discuss the derivation of the macroscopic models starting from our proposed CA models.
In Sec. \ref{sec:KMC}, we describe the list-based KMC algorithm and its implementation.
In Sec. \ref{sec:numerics}, we provide a series of numerical simulations in various parameter regimes for the 1D flows, and compare the microscopic and macroscopic models. Finally, we state our conclusions in Sec. \ref{sec:conclusion}.

\section{Cellular Automata Models with Look-ahead Rules}\label{sec:micro}

We describe the construction of the cellular automata (CA) model for 1D traffic flow in this section. The CA model is defined on a periodic lattice $\mathcal{L}$ with $\NL$ evenly spaced cells, $\mathcal{L}=\{1,2,\ldots,\NL\}$. For simplicity,  we assume that all cars move toward one direction on a single-lane loop highway with no entrances or exits. The configuration at each cell $i\in \mathcal{L}$ is defined by an index $\sigma_i$ :
\begin{equation}
\sigma_i=
\begin{cases}
    1 \qquad \mbox{if a car occupies cell $i$}, \\
    0 \qquad \mbox{if the cell $i$ is empty}.
   \end{cases}                                       \label{eq:sigma}
\end{equation}
The state of the system is represented by $\sigma=\{\sigma_i\}_{i=1}^{\NL}$, which lies in the configuration space $\Sigma=\{0,1\}^\NL$.

\subsection{Interaction rules}\label{subsec:SK}
We now describe the dynamics of the CA models.
The car movement can be represented by the transitions in the state of the system, which obey the rules of an exclusion process \cite{Lig85}: two nearest-neighbor lattice cells exchange values in each transition and cars cannot occupy the same cell.
In addition, cars are only allowed to move one cell to the right in one transition. Therefore, the only possible configuration changes are of the form
\begin{equation}\label{eq:jump1}
\{\sigma_i=1, \sigma_{i+1}=0\} \rightarrow \{\sigma_i=0, \sigma_{i+1}=1\}.
\end{equation}

The transition rate for \eqref{eq:jump1} depends on spatial Arrhenius type one-sided interactions and a look-ahead feature to represent drivers' behavior. These rules allow cars (or drivers) to perceive the traffic situation up to $\L$ cells ahead in which $\L$ is the look-ahead range parameter. The interactions between a pair of successive cars cannot be neglected if the gap between them is shorter than $\L$; in such situations, the following car must decelerate to avoid collision with the leading car. This is similar to the spin-exchange Arrhenius dynamics in which the simulation is driven based on the energy barrier a particle has to overcome in changing from one state to another \cite{SKa06, AlS08}. During a spin-exchange between nearest-neighbor sites $i$ and $i+1$, the system will allow the index $\sigma_i$ at location $i$ to exchange its index with the one at $i+1$. This is interpreted as a car move from $i$ to $i+1$ with the rate given by the Arrhenius relation:
\begin{equation}
    r_i=\omega_0\exp\big(-\Eb(i)\big),                         \label{eq:rate1}
\end{equation}
where the prefactor $\omega_0=1/\tau_0$ corresponds to the car moving frequency or speed and $\tau_0$ is the characteristic time. The moving energy barrier $\Eb(i)$ for the car located in the $i$-th cell describes the slow-down effect due to the traffic in front. It is assumed to depend only on the traffic situation up to range $\L$ ahead of the car under consideration, namely $\Eb(i)$ depends on $\{\sigma_j\}_{j=i+2}^{i+L}$.

\begin{figure}[ht]
\begin{tikzpicture}
  \node at (3,5.3) {$(a)$};
  \node at (.5,-.4) {\footnotesize cell $i$};
  \foreach[evaluate={\z=int(4-\y)}] \y in {0,...,4}{
   \draw[pattern=north west lines, pattern color=blue!30] (1,\y) rectangle (5-\y,\y+.7);
   \draw[thick] (-.5,\y) -- (5.5,\y);
   \draw[thick] (-.5,\y+.7) -- (5.5,\y+.7);
   \foreach \x in {0,...,5}{ \draw[thick] (\x,\y) -- (\x,\y+.7); }
   \node at (6.5,\y+.35) {$\Nv=\z$};
   \draw[very thick, fill=gray!20] (.5, \y+.35) circle (.2);
   \ifthenelse{\y=0}{}{\draw[very thick, fill=blue] (\z+1.5, \y+.35) circle (.2);}
   \ifthenelse{\y>1}{\draw[very thick, fill=black] (4.5, \y+.35) circle (.2);}{}
  }
  \draw[very thick, fill=black] (2.5, 4.35) circle (.2);
  \draw[thick,<-] (1.1, -.4) -- (2.7, -.4);
  \draw[thick,->] (3.3, -.4) -- (4.9, -.4);
  \node at (3, -.4) {$\L$};
  \draw[thick,dashed] (1,-.1) -- (1,-.7);
  \draw[thick,dashed] (5,.-.1) -- (5,-.7);
\end{tikzpicture}
\hfill
\begin{tikzpicture}
  \node at (3,5.3) {$(b)$};
  \node at (.5,-.4) {\footnotesize cell $i$};
  \draw[pattern=north west lines, pattern color=blue!30] (2,4) rectangle (3,4.7);
  \draw[very thick, fill=blue] (2.5, 4.35) circle (.2);
  \draw[pattern=north west lines, pattern color=blue!30] (3,1) rectangle (4,1.7);
  \draw[very thick, fill=blue] (3.5, 1.35) circle (.2);
  \foreach[evaluate={\z=int(\y-1)}] \y in {0,...,4}{
   \ifthenelse{\y>1}{
     \draw[pattern=north west lines, pattern color=blue!30] (4,\y) rectangle (5,\y+.7);
     \draw[pattern=north west lines, pattern color=blue!30] (5-\y,\y) rectangle (6-\y,\y+.7);
     \draw[very thick, fill=blue] (4.5, \y+.35) circle (.2);
     \draw[very thick, fill=blue] (5.5-\y, \y+.35) circle (.2);
   }{}
   \draw[thick] (-.5,\y) -- (5.5,\y);
   \draw[thick] (-.5,\y+.7) -- (5.5,\y+.7);
   \foreach \x in {0,...,5}{ \draw[thick] (\x,\y) -- (\x,\y+.7); }
   \node at (6.5,\y+.35) {$\Nc=\ifthenelse{\y<3}{\y}{\z}$};
   \draw[very thick, fill=gray!20] (.5, \y+.35) circle (.2);
  }
  \draw[thick,<-] (1.1, -.4) -- (2.7, -.4);
  \draw[thick,->] (3.3, -.4) -- (4.9, -.4);
  \node at (3, -.4) {$\L$};
  \draw[thick,dashed] (1,-.1) -- (1,-.7);
  \draw[thick,dashed] (5,.-.1) -- (5,-.7);
\end{tikzpicture}
\caption{Schematic representation of two look-ahead rules. (a): The rule based on the distance $\Nv$: a car in gray with different numbers ($\Nv$) of vacant cells between it and the first car (in blue) ahead of it in the range $\L$  (here, $\L=4$). (b): The rule based on the density $\Nc/\L$: a car in gray with different numbers ($\Nc$) of cars (in blue) ahead of it in the range $\L$.}
                                                           \label{f1.car}
\end{figure}
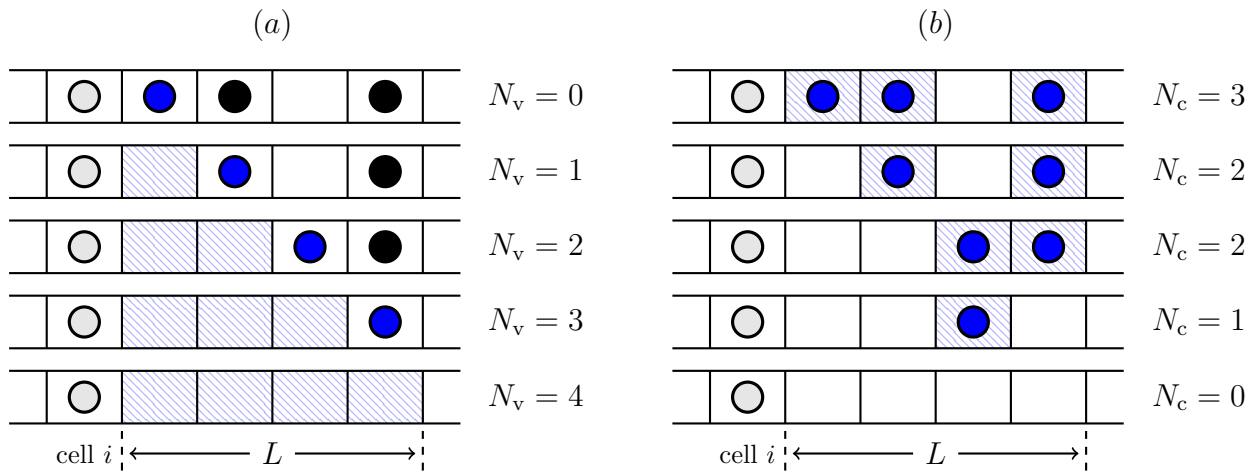

The energy barrier can be modeled through $\Eb = E_{\rm s} + E_{\rm c}$, where $E_{\rm s}$ is the external potential associated with the site binding of the car, which could vary in both space and time to account for spatial and temporal traffic situations, such as rush hour traffic, local weather anomalies, etc. In this study, we set $E_{\rm s}=0$. The term $E_{\rm c}$ enforces the look-ahead rules (Fig.~\ref{f1.car}). We consider the two rules in \cite{SuT14}, described as follows.

The first look-ahead rule is based on the distance from the car under consideration to the car in front of it, in other words, the number of vacant cells,  $\Nv$, between these two cars (as shown in Fig.~\ref{f1.car}(a)). Therefore, the energy barrier is given by
\begin{equation}
    \Eb(i) = \frac{\L-\Nv(i)}{\L}\Eo,   \label{eq:Eb1}
\end{equation}
where the parameter $\Eo$ is the car look-ahead interaction strength.
Based on the formulas  \eqref{eq:rate1} and \eqref{eq:Eb1}, we can see that the smaller is the value of $\Nv$, the larger is the energy barrier $\Eb$, thus the smaller is the transition rate $r$. This reflects the fact that the closer is the distance between cars, the stronger is the slowdown factor.

The second rule is based on the density of cars ahead of the car under consideration \cite{SKa06}. This is due to the fact that in real traffic drivers usually observe not only the leading car but also other cars ahead of the leading car.  In this rule, the energy barrier is given by
\begin{equation}
    \Eb(i) = \frac{\Nc(i)}{\L}\Eo,   \label{eq:Eb2}
\end{equation}
where $\Nc$ is the number of cars in the range $\L$ ahead of the car under consideration (as shown in Fig.~\ref{f1.car}(b)). It can be defined as
\[\Nc(i)=\sum_{j=i+1}^{i+L}\sigma_j.\]
This look-ahead rule with the formula \eqref{eq:Eb2} indicates that a slowdown factor is stronger when the forward car density is high, i.e., when the road is congested.

The coarse-grained macroscopic dynamics corresponding to the two CA models above are LWR model \eqref{eq:LWR} and SK model \eqref{eq:SK}, respectively. The formal derivation for the latter case can be found, for instance, in \cite{HST14,SKa06}.

\subsection{A new class of models}
We propose a new class of CA models, and show their relations to the macroscopic models with a non-concave flux \eqref{eq:fluxcc}, for any parameter $\Jump\in\mathbb{Z}_+$.

A novel discovery is that the parameter $\Jump$ in the macroscopic dynamics is related to the number of steps a car moves in a transition of the microscopic states.
A car can make multiple $\Jump$-moves, if the $\Jump$ cells in front are not occupied.

The new rule only allows the following configuration change
\begin{equation}\label{eq:jump}
\{\sigma_i=1, \sigma_{i+1}=\cdots=\sigma_{i+\Jump}=0\} \rightarrow
\{\sigma_i=\cdots=\sigma_{i+\Jump-1}=0, \sigma_{i+\Jump}=1\}.
\end{equation}
It represents a car move from $i$ to $i+\Jump$. The transition rate is modeled similarly as \eqref{eq:rate1}. Since the car moves $\Jump$ cells, we take the following rate
\begin{equation}
    r_i=\frac{\omega_0}{\Jump}\exp\big(-\Eb(i)\big), \label{eq:rate}
\end{equation}
so that the estimated velocity is comparable among different choices of $\Jump$.

Note that the rule \eqref{eq:jump} only allows cars to move if no cars occupy the $\Jump$ cells in front. Therefore, heuristically speaking,
the larger $\Jump$ is, the slower the cars move.
Such a phenomenon is also shared by the macroscopic dynamics \eqref{eq:main}, which is verified in the numerical experiments in Sec. \ref{subsect.jump}.

\section{Coarse-grained macroscopic models}\label{sec:macro}

In this section, we perform a formal derivation of the coarse-grained macroscopic model for our CA model.

\subsection{Semi-discrete mesoscopic models}
In a time step $\Delta\tt$, the probability of the configuration change
\begin{equation}\label{eq:prob}
  \mathbb{P}\big(\{\sigma_i=1, \sigma_{i+1}=\cdots=\sigma_{i+\Jump}=0\} \rightarrow
  \{\sigma_i=\cdots=\sigma_{i+\Jump-1}=0, \sigma_{i+\Jump}=1\}\big)
  =(\Delta\tt)~ r_i,
\end{equation}
where the rate $r_i$ is given in \eqref{eq:rate}.

Followed from \cite{HST14,SKa06}, we define $\sigma(\tt)=\{\sigma_i(\tt)\}_{i=1}^{\NL}$ be
a continuous-in-time stochastic process with a generator
\begin{equation}\label{eq:generator}
(A\psi)(\tt)=\lim_{\Delta\tt\to0}\frac{\E[\psi(\sigma(\tt+\Delta\tt))]
-\psi(\sigma(\tt))}{\Delta\tt},
\end{equation}
for any test function $\psi:\Sigma\to\mathbb{R}$, where $\tt$ is the time variable. All possible configuration changes from $\sigma(\tt)$ to $\sigma(\tt+\Delta\tt)$ obey the transition rule \eqref{eq:prob}.

By the definition of the generator, we have
\begin{equation}\label{eq:Egen}
  \frac{d}{d\tt}\E\psi=\E[A\psi].
\end{equation}

In particular, let us take $\psi(\sigma)=\sigma_i$. We calculate \eqref{eq:generator}
explicitly, and obtain
\begin{equation}\label{eq:Asigma}
A\sigma_i(\tt)=-r_i(\tt)\sigma_i(\tt)\prod_{j=1}^\Jump(1-\sigma_{i+j}(\tt))
+r_{i-\Jump}(\tt)\sigma_{i-J}(\tt)\prod_{j=1}^\Jump (1-\sigma_{i-\Jump+j}(\tt))
=: F_{i-\Jump}(\tt)-F_i(\tt),
\end{equation}
where $F_i$ is defined as
\[F_i(\tt)=r_i(\tt)\sigma_i(\tt)\prod_{j=1}^\Jump(1-\sigma_{i+j}(\tt)).\]

Let $\rho_i(\tt)=\E[\sigma_i(\tt)]=\mathbb{P}(\sigma_i(\tt)=1)$. Then, from \eqref{eq:Egen} and \eqref{eq:Asigma}, the dynamics of $\{\rho_i\}_{i=1}^{\NL}$ reads
\begin{equation}\label{eq:semirho}
  \frac{d}{d\tt}\rho_i(\tt)=\E[A\sigma_i(\tt)]=\E[F_{i-J}(\tt)]-\E[F_i(\tt)].
\end{equation}
Note that the right hand side of the equation is not yet a closed form of $\{\rho_i(\tt)\}_{i=1}^{\NL}$. We shall approximate the term $\E[F_i(\tt)]$ and make a closure to the system.

\subsection{Approximations as $\NL\to\infty$}
We start with a crucial assumption that helps us to obtain a closed system. It is called the \emph{propagation of chaos}, which means that $\{\sigma_i(\tt)\}_{i=1}^{\NL}$ are independent to each other, namely
\begin{equation}\label{eq:chaos}
  \E[\sigma_i(\tt)\sigma_j(\tt)]=\E[\sigma_i(\tt)]~\E[\sigma_j(\tt)],\quad \forall~i\neq j,~~t\geq0.
\end{equation}
Due to the look-ahead interaction, condition \eqref{eq:chaos} is not true for a system with fixed $\NL$ cells. However, as the number of cells $\NL$ tends to infinity, the system can become chaotic, and condition \eqref{eq:chaos} can be valid as $\NL\to\infty$.

By formally assuming the chaotic condition \eqref{eq:chaos}, we get
\[\E[F_i(\tt)]= \rho_i(\tt)\prod_{j=1}^J(1-\rho_{i+j}(\tt))~ \E[~r_i(\tt)~|~\sigma_i(\tt)=1,\sigma_{i+1}(\tt)=\cdots=\sigma_{i+\Jump}(\tt)=0~].
\]
For the rest of the section, we drop the $\tt$-dependence for simplicity.

To estimate the rate $r_i=\frac{\omega_0}{\Jump}\exp(-\Eb(i))$, we perform a formal Taylor expansion on $\Eb(i)$ around its mean $\E[\Eb(i)]$.
\[e^{-\Eb(i)}=e^{-\E[\Eb(i)]}\sum_{n=0}^\infty\frac{(-1)^n}{n!}\big(\Eb(i)-\E[\Eb(i)]\big)^n.\]
Taking the expectation, we obtain
\begin{equation}\label{eq:Eri}
  \E[r_i]=\frac{\omega_0}{\Jump} e^{-\E[\Eb(i)]}\left(1+\sum_{n=2}^\infty\frac{(-1)^n}{n!}\E\big[\big(\Eb(i)-\E[\Eb(i)]\big)^n\big]\right).
\end{equation}

Next, we estimate the energy barrier $\Eb$. To proceed, we first introduce the relative look-ahead distance $\Ld$, defined as
\begin{equation}\label{eq:macroL}
  \Ld = \frac{\L}{\NL}.
\end{equation}
We assume $\Ld$ is a fixed positive number ($0<\Ld\leq1$). So when $\NL\to\infty$, the look-ahead distance $\L=\Ld\NL$ would also tends to infinity.

The two barriers \eqref{eq:Eb1} and \eqref{eq:Eb2} will be discussed separately.

Recall the first energy barrier \eqref{eq:Eb1}
\[\Eb(i)=\frac{\L-\Nv(i)}{\L}\Eo,\]
where the random variable $\Nv(i)$ takes integer values between $0$ and $\L$. Conditioned with the configuration
$\{\sigma_i=1,\sigma_{i+1}=\cdots=\sigma_{i+\Jump}=0\}$,
$\Nv(i)$ takes values in $\{\Jump, \Jump+1, \ldots, \L\}$.

We impose the following assumption
\begin{equation}\label{eq:Nvscaling}
  \max \Nv(i)\ll\L=\Ld\NL,
\end{equation}
where the maximum is taken across all possible values of $\Nv$ and all locations $i=1,\ldots,\NL$.

The assumption \eqref{eq:Nvscaling} describes the scenario that
many cars are on the road, so that no two neighboring cars have a large distance comparable to the look-ahead distance $\L$.

Under the assumption \eqref{eq:Nvscaling}, we get
\[\E[\Eb(i)]=\frac{\L-\E[\Nv(i)]}{L}\Eo=\left(1-\frac{\E[\Nv(i)]}{L}\right)\Eo
  \xrightarrow{~\L\to\infty~}\Eo,\]
and consequently,
\[\Eb(i)-\E[\Eb(i)]=\frac{\L-\Nv(i)}{L}\Eo-\Eo=-\frac{\Nv(i)}{L}\Eo
  \xrightarrow{~\L\to\infty~}0.\]
Therefore, plug back into \eqref{eq:Eri}, we end up with
\[\E[r_i]\xrightarrow{~\L\to\infty~}\frac{\omega_0}{\Jump}e^{-\Eo}.\]

Next, we move to the second energy barrier \eqref{eq:Eb2}
\[\Eb(i)=\frac{\Eo}{\L}\sum_{j=i+1}^{i+\L}\sigma_j
  =\frac{\Eo}{\L}\sum_{j=i+J+1}^{i+\L}\sigma_j,\]
where the second equality is valid under the conditional configuration
$\{\sigma_i(\tt)=1,\sigma_{i+1}(\tt)=\cdots=\sigma_{i+\Jump}(\tt)=0\}$.
Compute
\[\E[\Eb(i)]=\frac{\Eo}{\L}\sum_{j=i+J+1}^{i+\L}\rho_j.\]
Then, we have
\begin{align*} &\E[(\Eb(i)-\E[\Eb(i)])^2]=\frac{\Eo^2}{\L^2}\E\left[\left(\sum_{j=i+J+1}^{i+\L}(\sigma_j-\rho_j)\right)^2\right]\\&                                                                                                  =\frac{\Eo^2}{\L^2}\E\left[\sum_{j=i+J+1}^{i+\L}(\sigma_j-\rho_j)^2\right]+
\frac{2\Eo^2}{\L^2}\E\left[\sum_{j=i+J+1}^{i+\L}\sum_{k=i+J+1}^{j-1}(\sigma_j-\rho_j)(\sigma_k-\rho_k)\right].
\end{align*}
By condition \eqref{eq:chaos}, the cross terms
\[\E[(\sigma_j-\rho_j)(\sigma_k-\rho_k)]=\E[\sigma_j-\rho_j]\E[\sigma_k-\rho_k]=0.\]
Moreover, since $|\sigma_j-\rho_j|\leq1$, we have $\E[(\sigma_j-\rho_j)^2]\leq 1$. Hence,
\[\E[(\Eb(i)-\E[\Eb(i)])^2]=\frac{\Eo^2}{\L^2}\sum_{j=i+J+1}^{i+\L}\E[(\sigma_j-\rho_j)^2]\leq \frac{\Eo^2(\L-\Jump)}{\L^2}\xrightarrow{~\L\to\infty~}0.\]
Similarly, higher moments vanishes when $\L\to\infty$:
\[\E[(\Eb(i)-\E[\Eb(i)])^n]\xrightarrow{~\L\to\infty~}0,\quad\forall~n\geq2.\]
Plug back into \eqref{eq:Eri}, we conclude with
\[\E[r_i]\xrightarrow{~\L\to\infty~}\frac{\omega_0}{\Jump}\exp\left(-\frac{\Eo}{\L}\sum_{j=i+\Jump+1}^{i+\L}\rho_j\right).\]

To sum up, we achieve an approximation of $\E[F_i(\tt)]$ in terms of $\{\rho_i(\tt)\}_{i=1}^\NL$ as $\NL\to\infty$
\begin{equation}\label{eq:fluxlimit}
  \E[F_i(\tt)]\xrightarrow{~\NL\to\infty~}\begin{cases}
   \displaystyle\frac{\omega_0}{\Jump}\rho_i(\tt)\prod_{j=1}^J(1-\rho_{i+j}(\tt))e^{-\Eo}
    & \text{First rule}\\ \displaystyle\frac{\omega_0}{\Jump}\rho_i(\tt)\prod_{j=1}^J(1-\rho_{i+j}(\tt))\exp\left(-\frac{\Eo}{\L}\sum_{j=i+\Jump+1}^{i+\L}\rho_j(\tt)\right)  & \text{Second rule}
  \end{cases}
\end{equation}

\subsection{Coarse-grained PDE models}
We rescale the lattice $\mathcal{L}$ into a fixed interval $\Omega=[0,1]$, where each cell has length $h=1/\NL$.  The $i$-th cell is rescaled to the interval
$[(i-1)h, ih]$.

Define the macroscopic density
$\rho~:~\Omega\times\mathbb{R}_+\to\mathbb{R}$, where
\[\rho(x,\t) = \rho_i(\tt),\quad\text{with }~x=ih,~~\t= \tt h.\]

A coarse-grained model can be obtained by formally letting $h\to0$.
Under our setup, the parameters are scaled as
\[\underbrace{1\leq\Jump\leq\Nv}_{\mathcal{O}(1)}\ll
\underbrace{\L\leq\NL}_{\mathcal{O}(h^{-1})},\]
where $\L$ and $\NL$ goes to infinity with a fixed ratio $\Ld=\L/\NL$.

The flux in \eqref{eq:fluxlimit} as $h\to0$ has the form
\begin{numcases}{F(x,\t):=J\cdot\lim_{h\to0}\E[F_i(\tt)]=}
   \omega_0\rho(x,\t)(1-\rho(x,\t))^\Jump e^{-\Eo}, \label{eq:flux1}\\
   \omega_0\rho(x,\t)(1-\rho(x,\t))^\Jump\exp
   \left(-\Eo\displaystyle\int_x^{x+\Ld}\rho(y,\t)dy\right).
    \label{eq:flux2}
\end{numcases}
For the first rule, the macroscopic flux \eqref{eq:flux1} depends locally on the density $\rho$; while for the second rule, the flux \eqref{eq:flux2} is nonlocal in $\rho$.

The dynamics of $\rho$ in \eqref{eq:semirho} becomes the following scaler conservation law:
\begin{align*}
  \partial_\t\rho(x,\t)&=\frac{1}{h}\frac{d}{d\tt}\rho_i(\tt)
  =\frac{\E F_{i-\Jump}(\tt)-\E F_i(\tt)}{h}\\
&\xrightarrow{~h\to0~}\lim_{h\to0}\frac{F(x-\Jump h,\t)-F(x,\t)}{\Jump h}
=-\partial_x(F(x,\t)).
\end{align*}

We end up with the following coarse-grained PDE models:

(i). For \textbf{the first look-ahead rule},
\begin{equation}
    \partial_t\rho+\partial_x\left(\omega_0\rho(1-\rho)^\Jump e^{-\Eo}\right)=0.
    \label{eq:rule1}
  \end{equation}
This is the LWR type local model with flux \eqref{eq:fluxcc} and $u_{\max}=\omega_0e^{-\Eo}$.

(ii). For \textbf{the second look-ahead rule},
\begin{equation}
    \partial_t\rho+\partial_x\left(\omega_0\rho(1-\rho)^\Jump
      \exp\left(-\Eo\int_x^{x+\Ld}\rho(y,t)dy\right)\right)=0.
    \label{eq:rule2}
\end{equation}
This is indeed our proposed macroscopic model \eqref{eq:main} with a nonlocal look-ahead interaction kernel $K\equiv\Eo$.

If the relative look-ahead distance $\Ld=0$, the equation \eqref{eq:rule2} becomes the local dynamics \eqref{eq:rule1}. On the other hand, if we consider the periodic domain (loop highway) and set $\Ld=1$, namely $\L=\NL$, the interaction becomes global. By conservation of mass, the averaged car density
\[\int_x^{x+1}\rho(y,t)dy=\bar{\rho}\]
is a constant for any $x$ and $t$. Equation \eqref{eq:rule2} again reduces to the local dynamics \eqref{eq:rule1}, with a different $u_{\max}=\omega_0e^{-\bar{\rho}\Eo}$, which also depends on the constant $\bar{\rho}$.

\section{The Kinetic Monte Carlo Method} \label{sec:KMC}

To investigate the evolution of the nonlocal traffic system, we apply the kinetic Monte Carlo (KMC) method  \cite{BKL75}  to the microscopic CA model with look-ahead interactions. The reason to choose the KMC instead of the Metropolis Monte Carlo (MMC) method \cite{MRR53} is that trial steps in the MMC are sometimes rejected because the acceptance probability is small, in particular when a system approaches the equilibrium, or the density of cars is high. A main feature of the KMC algorithm is that it is ``rejection-free''. In each step, the transition rates for all possible changes from the current configuration are calculated and then a new configuration is chosen with a probability proportional to the rate of the corresponding transition. The other feature of the KMC method is its capability of providing a more accurate description of the real-time evolution of a traffic system in terms of these transition rates since the KMC method is more suitable for simulating the non-equilibrium system.

We emphasize that although the KMC algorithm was presented in \cite{SuT14}, there is a major update due to the multiple jumps $J$ introduced in the current work. To make the presentation  self-contained, we include the details of the algorithm here again. The KMC algorithm is built on the assumption that the model features $N$ independent Poisson processes (corresponding to $N$ moving cars on the lattice) with transition rates $r_i$ in \eqref{eq:rate} that sum up to give the total rate $R=\sum^{N}_{i=1}r_i$. In simulations with a finite number of distinct processes, it is more efficient to consider the groups of events according to their rates \cite{BBS95, Sch02, SCE11}. This can be done by forming lists of the same kinds of events according to the values of $\Nv$ in \eqref{eq:Eb1} of the first look-ahead rule or the values of $\Nc$ in \eqref{eq:Eb2} of the second look-ahead rule. Therefore, we can put the total $N$ events into $(\L+1)$ lists, labeled by $l=0, \ldots, \L$. All processes in the $l$-th list have the same rate $r_l$. We denote the number of processes in this list by $n_l$, which is called the \textit{multiplicity}, and we have $N=\sum^L_{l=0}n_l$. To each list, we assign a partial rate, $R_l = n_l r_l$, and a relative probability, $P_l = R_l/R$. Then the total rate is given by $R = \sum^L_{l=0} n_l r_l$. A fast list-based KMC algorithm at each KMC step based on the grouping of events is given as follows.

{\it List-based KMC algorithm:}
\label{al.bkl}

Step 1:  Generate a uniform random number, $\xi_1 \in (0, 1)$ and decide which  process will take place by choosing the list index $s$ such that
\begin{equation}
     \sum^{s-1}_{l=0} \frac{R_l}{R} < \xi_1 \leq \sum^s_{l=0} \frac{R_l}{R}
                                                           \label{eq.bkl}
\end{equation}

Step 2: Select a car for the realization of the process $s$. This can be done with the help of a list of coordinates for each kind of event, and an integer random number $\xi_2$ in the range $[1, n_s]$; $\xi_2$ is generated and the corresponding car/event from the list is selected.


Step 3: Check if there are enough vacant cells ahead of the selected car. If ``Yes" (i.e., $\Nv \geq \Jump$), perform Steps 4--6 for total $\Jump$ times so that the selected car can make $\Jump$ moves before continuing to the next KMC step. If ``No" (i.e., $\Nv < \Jump$), perform only Step 5 for total $\Jump$ times so that the selected car will stay in its current cell for $\Jump$ transition time periods before continuing to the next KMC step.

Step 4: Perform the selected event (the car move to the next cell) leading to a new configuration.

Step 5: Use $R$ and another random number $\xi_3 \in (0, 1)$ to decide the time it takes for that event to occur (the transition time), i.e., the nonuniform time step $\Delta t=-\log(\xi_3)/R$.

Step 6: Update the multiplicity $n_l$, relative rates $R_l$,
total rate $R$ and any data structure that may have changed due to this move. \hfill $\Box$

In summary, the following parameters need to be given for the KMC simulations with either look-ahead rule: (i) the characteristic time $\tau_0$; (ii) the car interaction strength $E_0$; (iii) the look-ahead parameter $\L$; and (iv) the multiple move parameter $\Jump$ ($1\leq \Jump \leq \L$).

\section{Numerical experiments}\label{sec:numerics}

We next investigate 1D nonlocal traffic flows in various parameter regimes with the numerical method presented in the previous section. We start by calibrating some KMC model parameters with respect to well-known quantities from real traffic data.

\subsection{Calibration and validity by the red light traffic problem}

Following \cite{SKa06,SuT14}, we set the actual physical length of each cell to $22$ feet ($\approx 6.7$m), which allows for the average car length plus safe distance. Therefore, $1$ mile ($=5280$ feet $\approx 1609$m) is equivalent to $240$ cells. For a car which has average speed of $60$ miles per hour ($\approx 26.8$ m/s), an estimate of time to cross a cell is given by
\begin{equation}
    \Delta \tt_{\rm cell}=\frac{22 \ {\rm feet}}{60 \ {\rm miles/h}}
    = \frac{1 \ {\rm cell} \times 3600 \ {\rm s}}{60 \times 240 \ {\rm cells}} = \frac{1}{4} {\rm s}.               \label{eq2.vel}
\end{equation}
We calibrate the parameters $\tau_0$ and $\Eo$ by simulating a free-flow regime where all cars are expected to drive at their desired speed that is set to $60$ miles per hour ($\approx 26.8$ m/s). This is accomplished by setting the characteristic time $\tau_0=0.25$s, and then $\omega_0=4$s$^{-1}$. In fact, due to the inherent stochasticity in the simulations, sometimes cars may move faster or slower than the speed limit. We also mention that other values of $\tau_0$ and $\Eo$ may be chosen to adjust our model for considering different standards in other regions or countries.

\begin{figure}[!ht]
\begin{center}
\includegraphics[width=.48\textwidth]{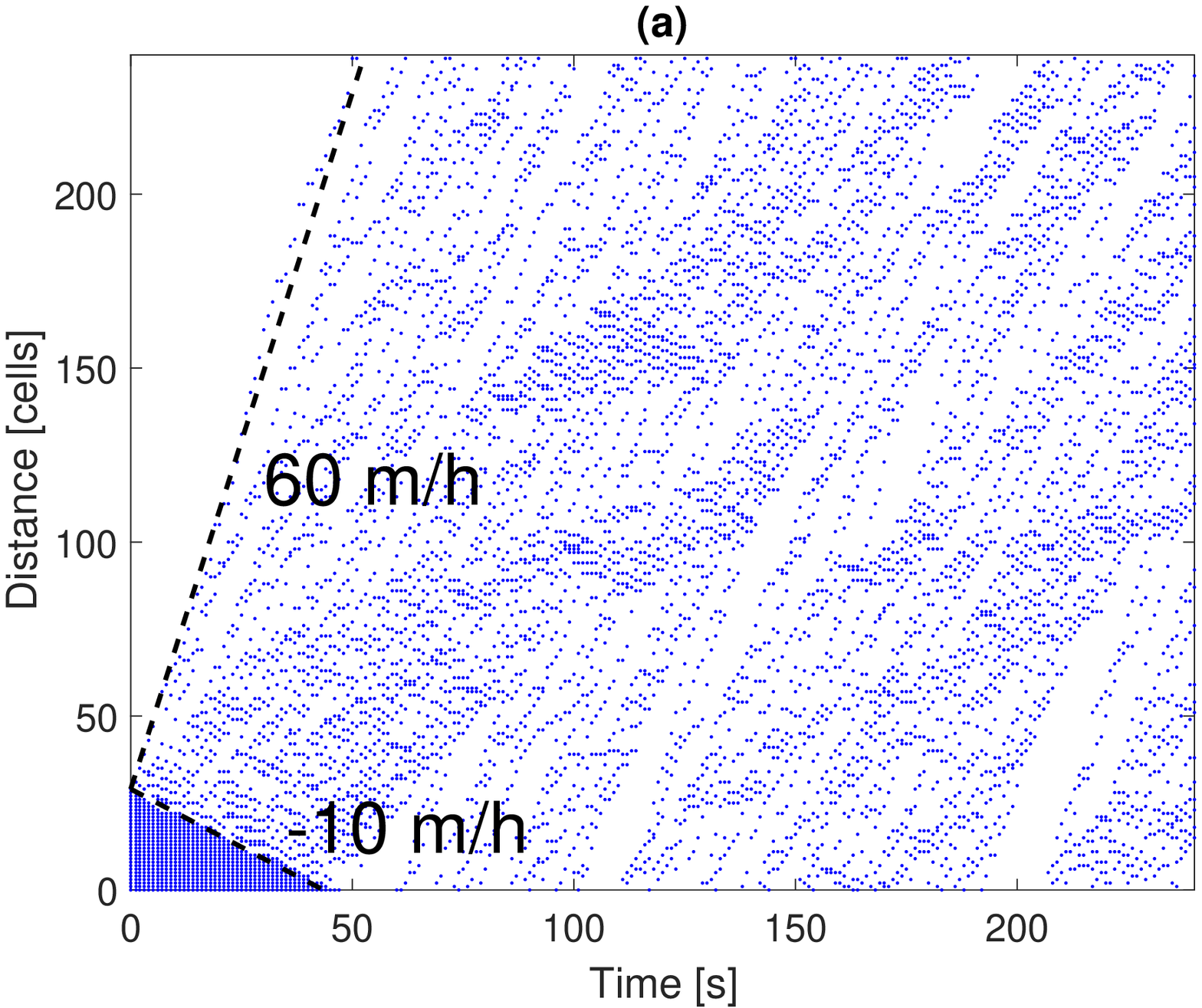} \hfill
\includegraphics[width=.48\textwidth]{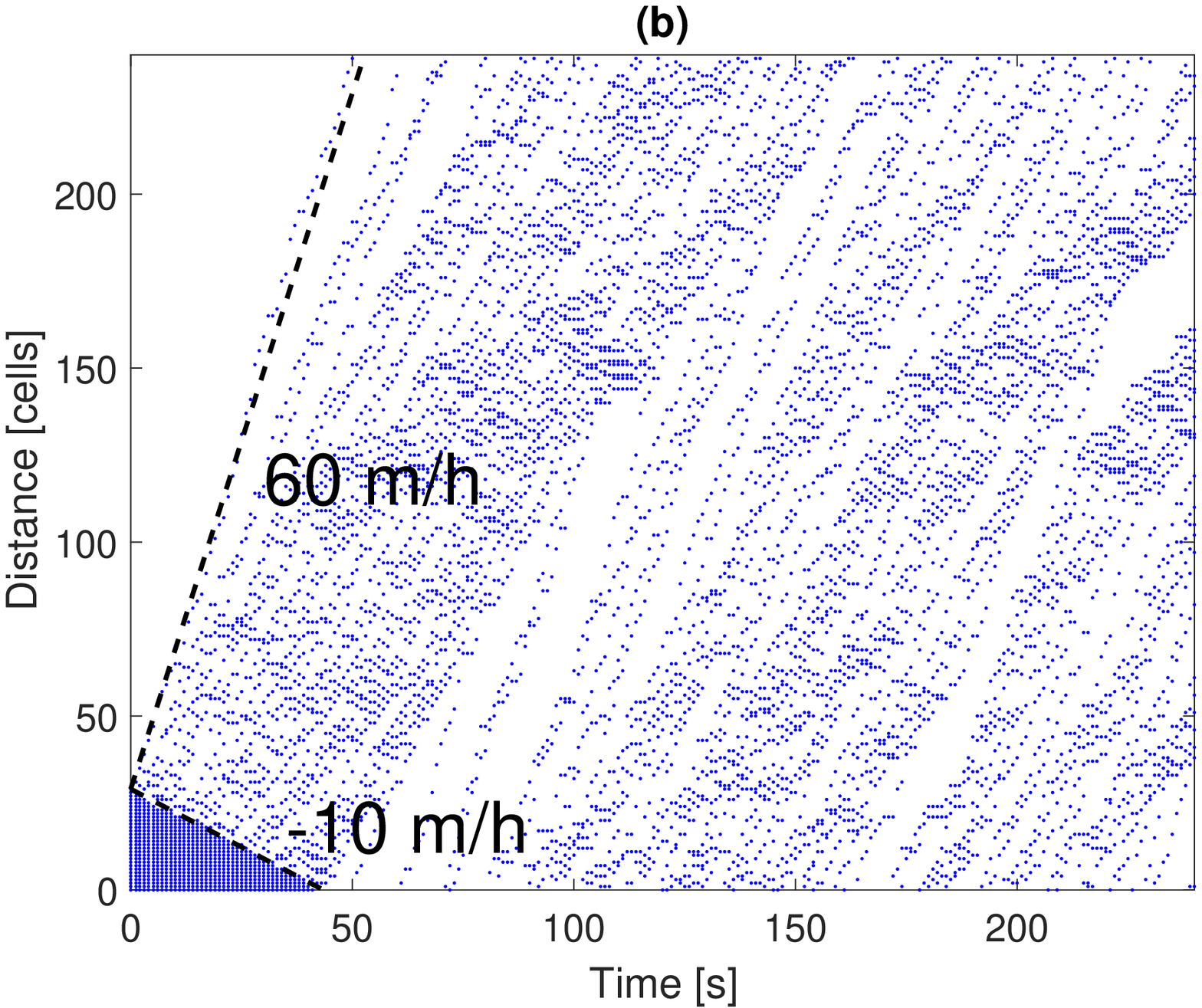}

\caption{Calibration of the interaction strength $\Eo$ permitting the desired car speed of $60$ miles per hour ($\approx 26.8$ m/s) and an upstream front velocity of $\approx-10$ miles per hour ($\approx -4.5$ m/s), which are indicated by the dashed lines, respectively. We take the highway distance of $1$ mile ($\approx 1609$ m, $\NL=240$ cells) and set the look-ahead parameter of $\L=4$ and the multiple move parameter of $\Jump=2$ for both look-ahead rules. The initial condition corresponds to a red light traffic problem, i.e., bumper-to-bumper cars up to $0.125$ miles ($\approx 201$ m $=30$ cells) and no cars after that, so a total of $30$ cars in each simulation. The running time is up to $240$ s.  (a)  Car traces in a simulation with the first look-ahead rule \eqref{eq:Eb1} and the interaction strength $\Eo=4.5$. (b): Car traces in a simulation with the second look-ahead rule \eqref{eq:Eb2} and $\Eo=6.0$.}
                                                           \label{fig:redlight}
\end{center}
\end{figure}

Fig.~\ref{fig:redlight} shows the results of the ``red light" traffic problem: the traffic light located at $0.125$ miles ($\approx 201$m, i.e., $i=30$ cells) is turned from red to green at the initial time and the ``bumper to bumper" traffic wave is released. The initial condition is given by
\begin{equation}
\sigma_i =
\begin{cases}
    1 \qquad 1 \leq i \leq 30, \\
    0 \qquad 31 \leq  i \leq \NL.
   \end{cases}                                       \label{eq2.init}
\end{equation}
The highway distance is set to $1$ mile ($\approx 1609$m), i.e., $\NL=240$ cells. Then the \emph{averaged car density} $\bar{\rho}=30/240=12.5\%$, which is in the free-flow regime. We calibrate the interaction strength $\Eo$ such that the velocity of an upstream front can be approximately $-10$ miles per hour ($\approx -4.5$ m/s), as estimated by traffic researchers in \cite{KeR97, HHS02, Scha02}. While we take the look-ahead parameter $\L=4$ and the multiple move parameter $\Jump=2$ for both look-ahead rules, the calibrated value of the interaction strength is $\Eo=4.5$ for the first look-ahead rule \eqref{eq:Eb1} (Fig.~\ref{fig:redlight} (a)) and $\Eo=6.0$ for the second look-ahead rule \eqref{eq:Eb2} (Fig.~\ref{fig:redlight}(b)).

\subsection{Numerical comparisons for different interaction strengths}
\label{subsect.strength}

\begin{figure}[!ht]
\begin{center}
\includegraphics[width=.48\textwidth]{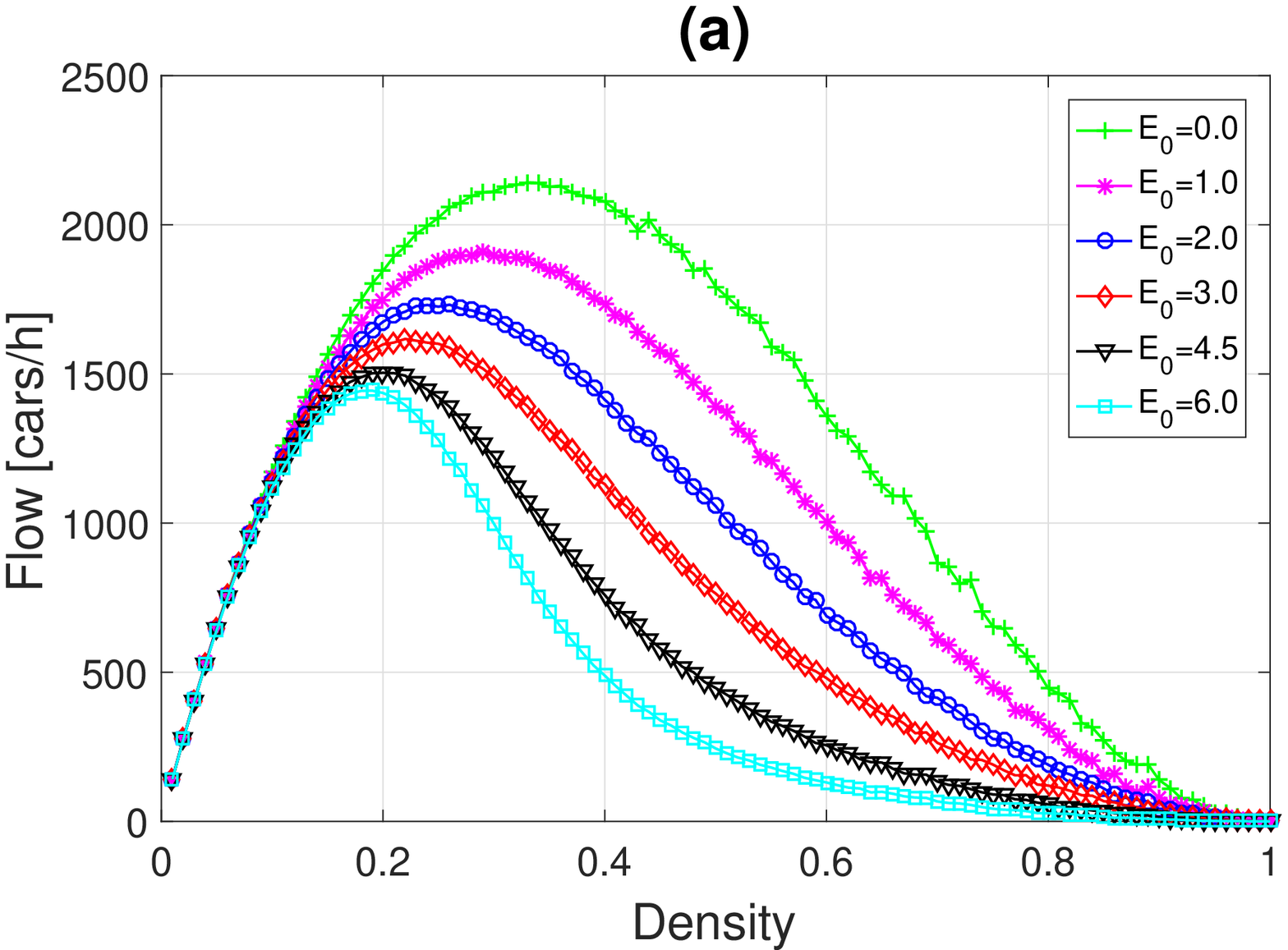} \hfill
\includegraphics[width=.48\textwidth]{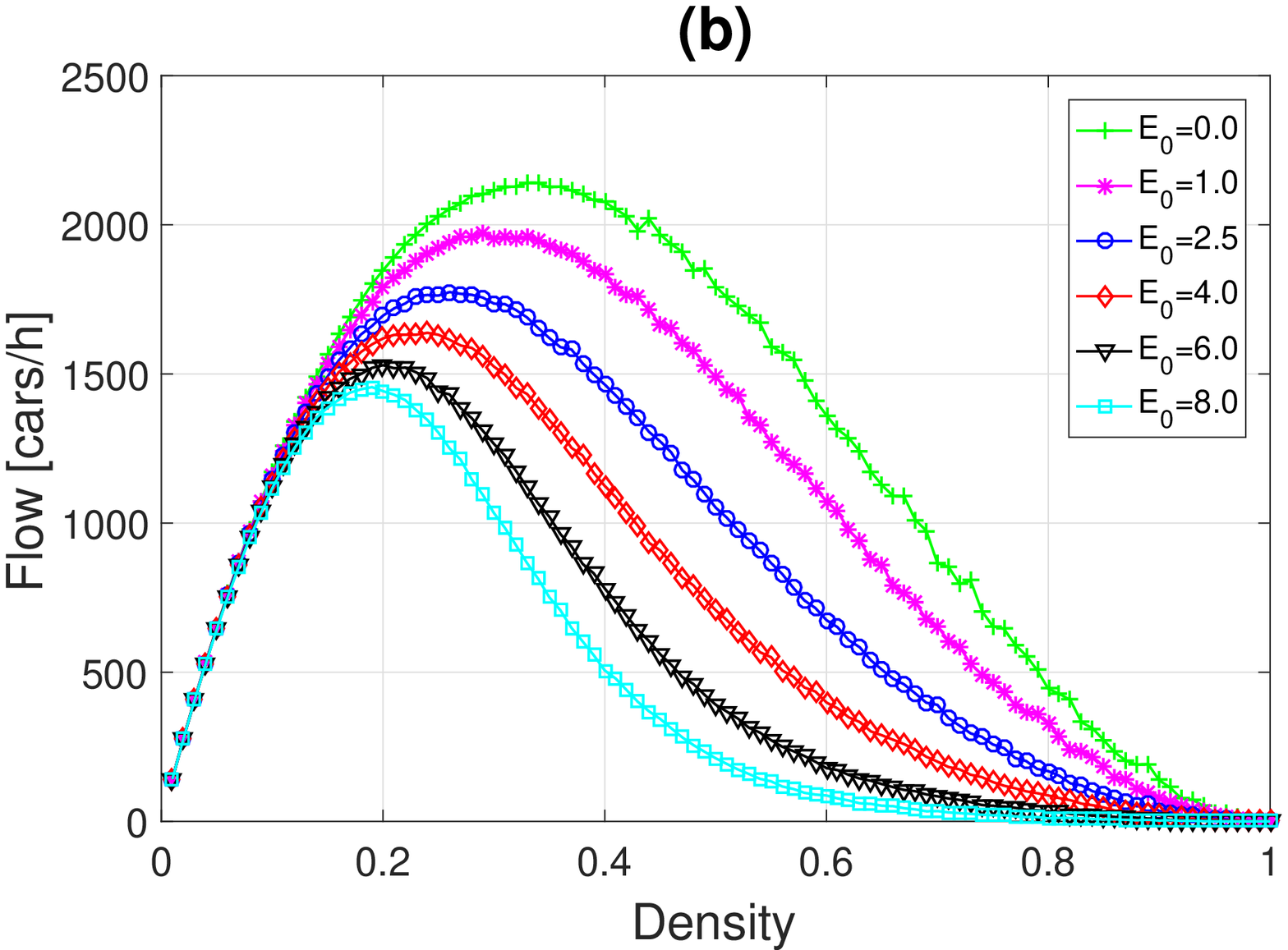}

\vspace{0.2cm}

\hspace{0.2cm} \includegraphics[width=.46\textwidth]{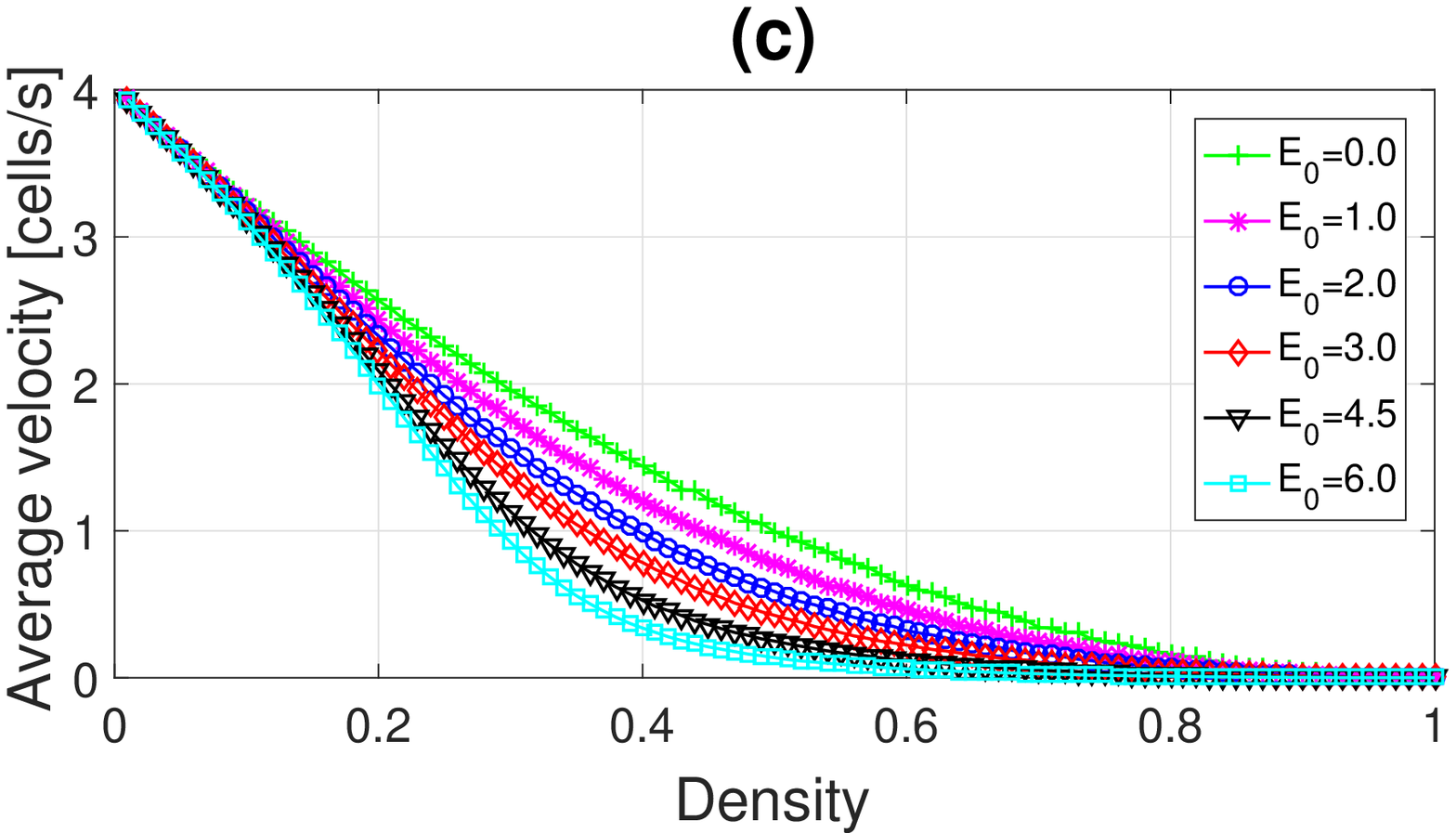} \hfill
\includegraphics[width=.46\textwidth]{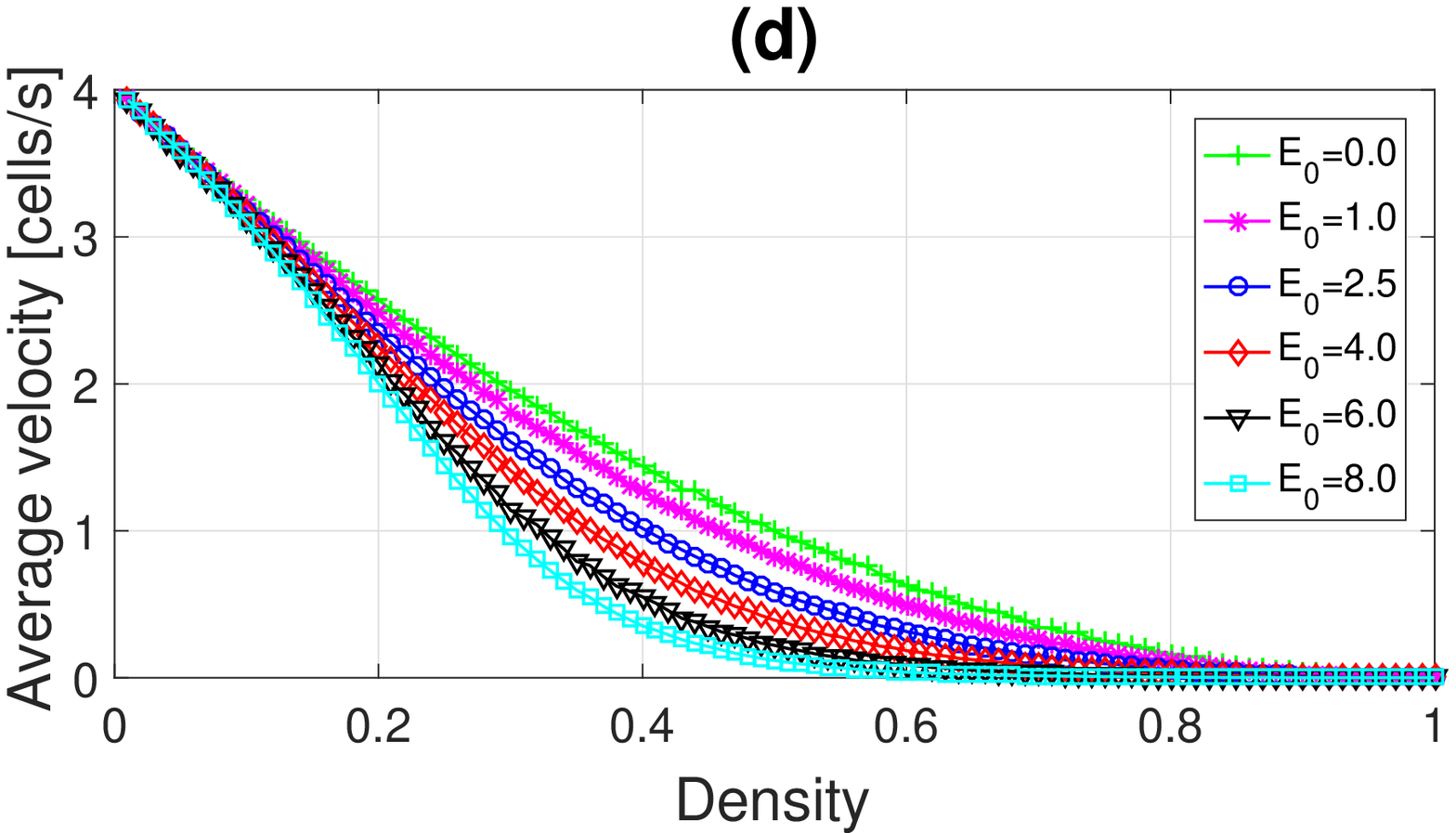}

\vspace{0.2cm}

\hspace{0.2cm} \includegraphics[width=.47\textwidth]{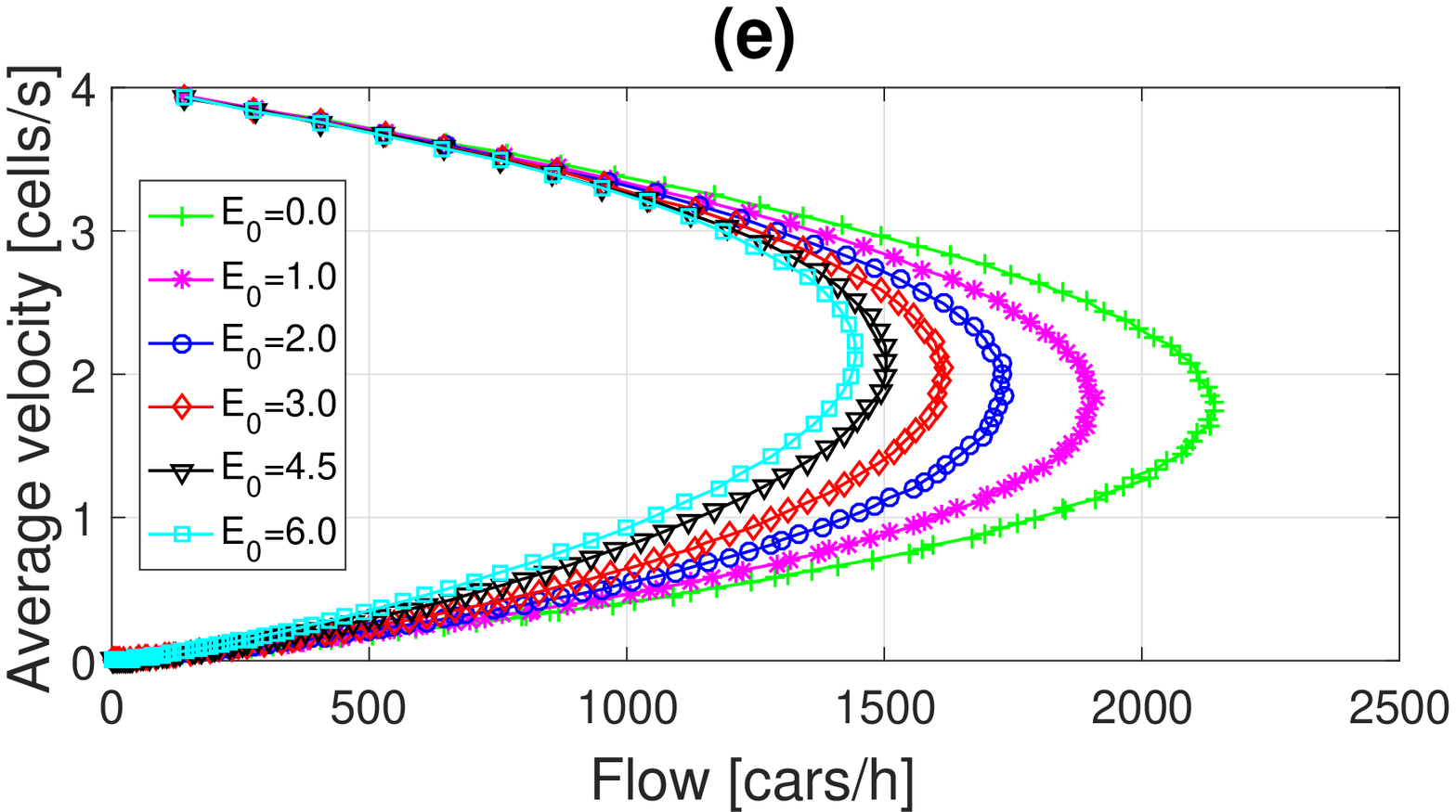} \hfill
\includegraphics[width=.47\textwidth]{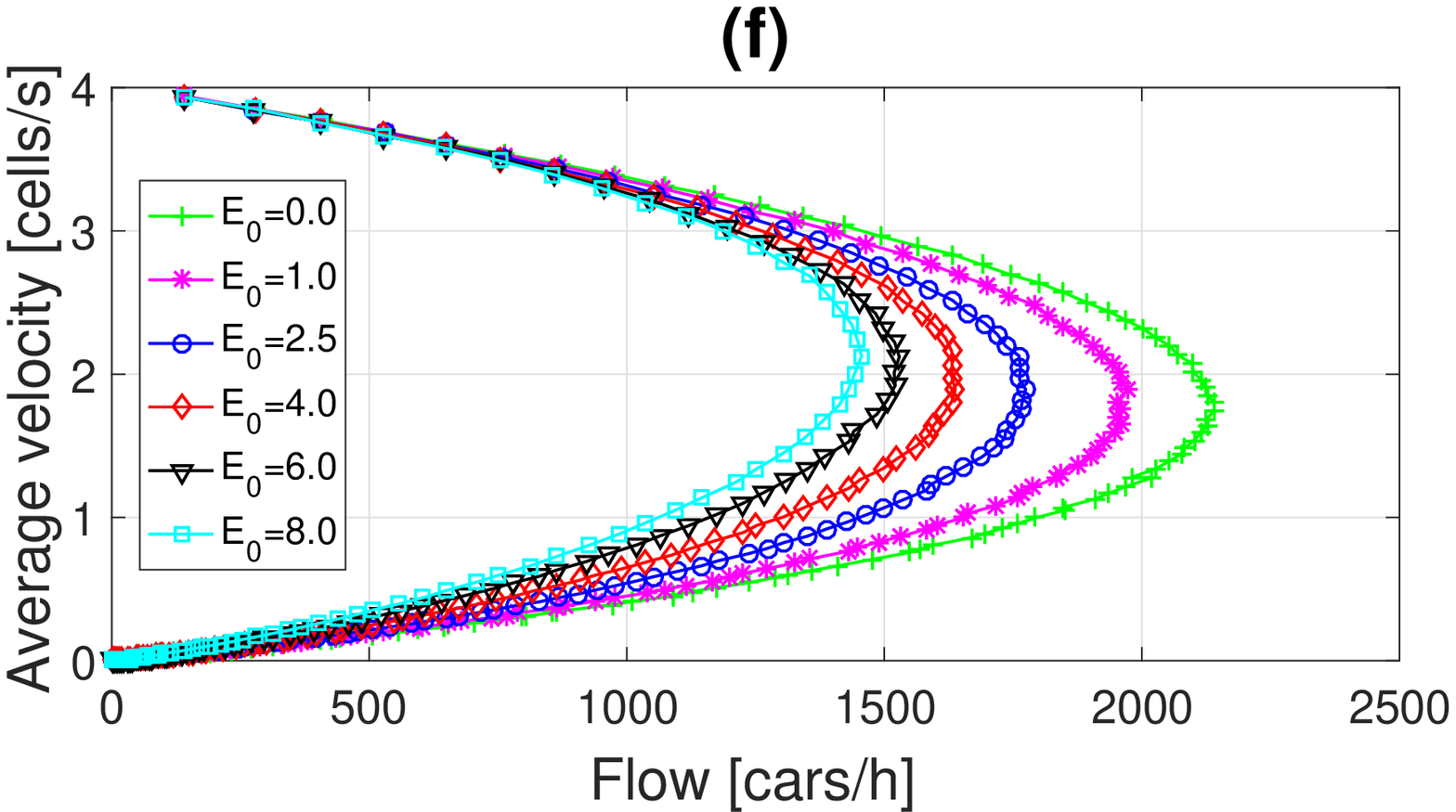}

\caption{Comparison results of the traffic flow on the one-lane highway with six different values of the interaction strength $\Eo$. In all KMC simulations, we take the highway distance of $\approx4.17$ miles ($\approx 6704$m, $\NL=1000$ cells), the look-ahead parameter of $\L=4$, the multiple move parameter of $\Jump=2$ and the final time of $1$ hour. (a)(b): Longtime averages of the density-flow relationship; (c)(d): Ensemble-averaged velocity of cars versus the density $\bar{\rho}$; (e)(f): Long-time averages of the flow-velocity relationship. (left panel): Results of the first look-ahead rule \eqref{eq:Eb1} with $\Eo=0$ to $6.0$. (right panel): Results of the second look-ahead rule \eqref{eq:Eb2} with $\Eo=0$ to $8.0$.}
                                                           \label{fig:flowE}
\end{center}
\end{figure}

First, we analyze the effects of the interaction strength $\Eo$ on the traffic flow and identify the range of significance of parameters. In the following we take a fixed look-ahead distance of $\L=4$ and the multiple move parameter $\Jump=2$ and make a series of numerical tests for different car densities with various values of parameters $\Eo$. We show the fundamental diagrams of the density-flow, density-velocity and flow-velocity relationships and compare the results of two look-ahead rules \eqref{eq:Eb1} and \eqref{eq:Eb2}. For these results we take a random car distribution at the initial time on a loop highway of $\approx4.17$ miles ($\approx 6704$m, $\NL=1000$ cells) and observe the behavior of traffic flows as the averaged car density $\bar{\rho}$ increases incrementally from $\bar{\rho}=0.01$ to $\bar{\rho}=0.99$. The traffic flow is measured as the number of cars passing a detector site per unit time \cite{May90}. In Fig.~\ref{fig:flowE}, we run each KMC simulation with different densities until the same final time ($1$ hour) and report long time averages of the flow $\Fbar$ in number of cars per hour and the ensemble-averaged velocity of all cars $\vbar$ in cells per second.

In Figs.~\ref{fig:flowE}(a) and (b), we plot the fundamental diagrams on the averaged fluxes $\Fbar$ against the averaged density $\bar\rho$ of the first look-ahead rule \eqref{eq:Eb1} with $\Eo=0$ to $6.0$ and the second look-ahead rule \eqref{eq:Eb2} with $\Eo=0$ to $8.0$, respectively.
They all share certain characteristics: a nearly linear increase of the flow at low averaged densities (which corresponds to the free-flow regime), a single maximum of the flow reached a critical density $\rhoc$, and a right-skewed asymmetry (namely $\rhoc<1/2$). The shape agrees with other simulation results and observational data \cite{NaS92, HHS02, LiT05}. We also observe that both the value of the critical density $\rhoc$ and the maximum value of the flow $\Fbar$ tend to decrease with increasing $\Eo$ because the larger is the interaction strength the stronger is the interaction to slow down the cars.


Figs.~\ref{fig:flowE}(c) and (d) show the fundamental diagrams of the density-velocity relationship for two look-ahead rules, respectively. In the free-flow regime the ensemble-averaged velocity $\vbar$ decreases approximately linearly from the maximum speed of $4$ cells per second ($\approx 26.8$ m/s or $60$ miles/h) as $\bar{\rho}$ increases and the chance of interaction between cars gets higher. As $\Eo$ increases, when $\bar{\rho}$ is larger than the critical point $\rhoc$, the average velocity drops down to zero and the density-velocity curve is negative exponential. This linear relationship follows the Greenshields model \cite{Gre35} and the negative exponential relationship belongs to the Underwood model \cite{Und61}.

Figs.~\ref{fig:flowE}(e) and (f) show the fundamental diagrams of the flow-velocity relationship for two look-ahead rules, respectively, which plot the ensemble-averaged velocity $\vbar$ versus the flow $\Fbar$. For the case of the interaction strength $\Eo=0$ (shown as green ``$+$" signs), the flow $\langle F \rangle$ reaches its maximum $\approx 2140$ cars per hour when the ensemble-averaged velocity $\vbar$ is at a critical value $\vbar_{\rm crit} \approx 1.7$ cells per second ($\approx 11.6$ m/s or $25.5$ miles/h). As $\Eo$ increases, the maximum value of the flow decreases and
the critical value $\vbar_{\rm crit}$ increases  and becomes higher than $2$ cells per second. The results compare favorably with observed data in \cite{Wie95}.

We remark that for a small look-ahead distance of $\L=4$, both rules produce similar results as shown in Fig.~\ref{fig:flowE}. Indeed, the macroscopic model \eqref{eq:rule1} is very close to \eqref{eq:rule2} with a small $\Ld=\frac{\L}{\NL}=0.004$.
We also note that in Figs.~\ref{fig:flowE}(a) and (b), the density-flow curves of the KMC simulations with $\Eo=4.5$ for the first look-ahead rule \eqref{eq:Eb1} and $\Eo=6.0$ for the second look-ahead rule \eqref{eq:Eb2} (shown as black ``$\bigtriangledown$" signs) clearly display that the region of free-flow persists up to the density of approximately $\rhoc =0.2$, i.e., $240\times0.2\approx50$ cars per mile. These results are naturally produced by the traffic dynamics in our simulations with the calibrated parameters $\tau_0=0.25$s, the look-ahead distance of $\L=4$ and the multiple move parameter $\Jump=2$, which agrees with observations \cite{Wie95, HHS02}.

\begin{figure}[!ht]
\begin{center}
\includegraphics[width=.48\textwidth]{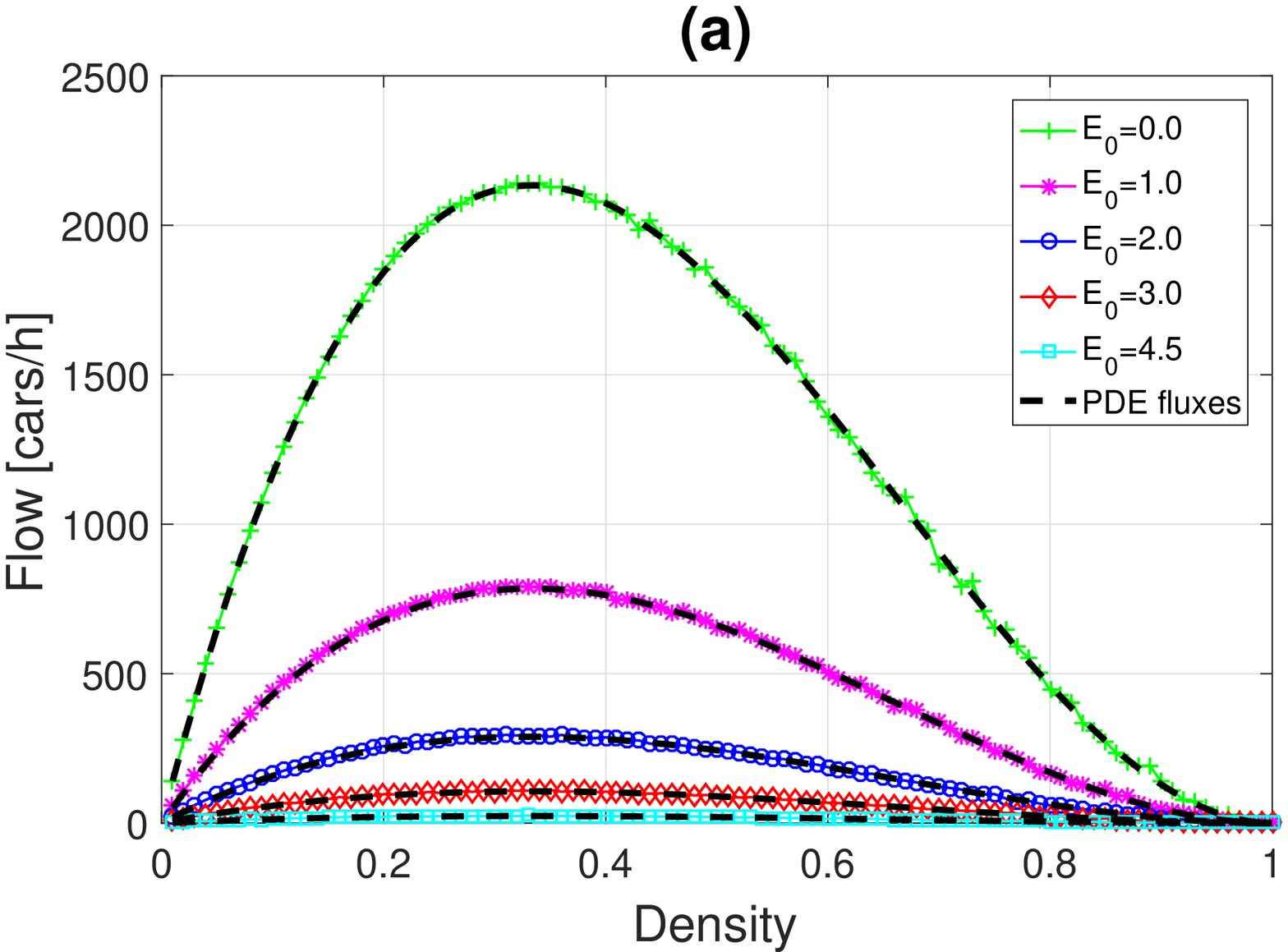} \hfill
\includegraphics[width=.48\textwidth]{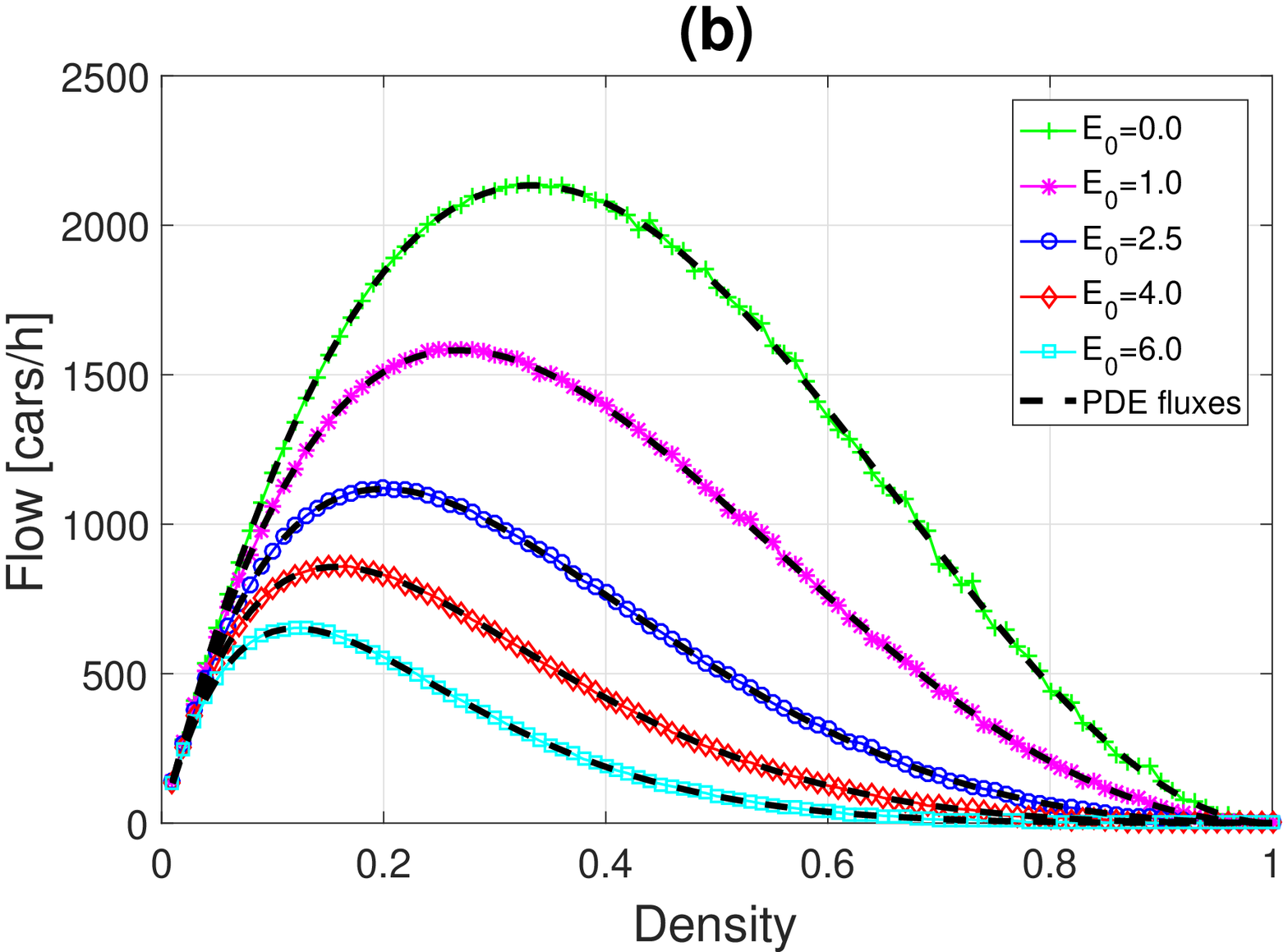}

\vspace{0.2cm}

\hspace{0.2cm} \includegraphics[width=.46\textwidth]{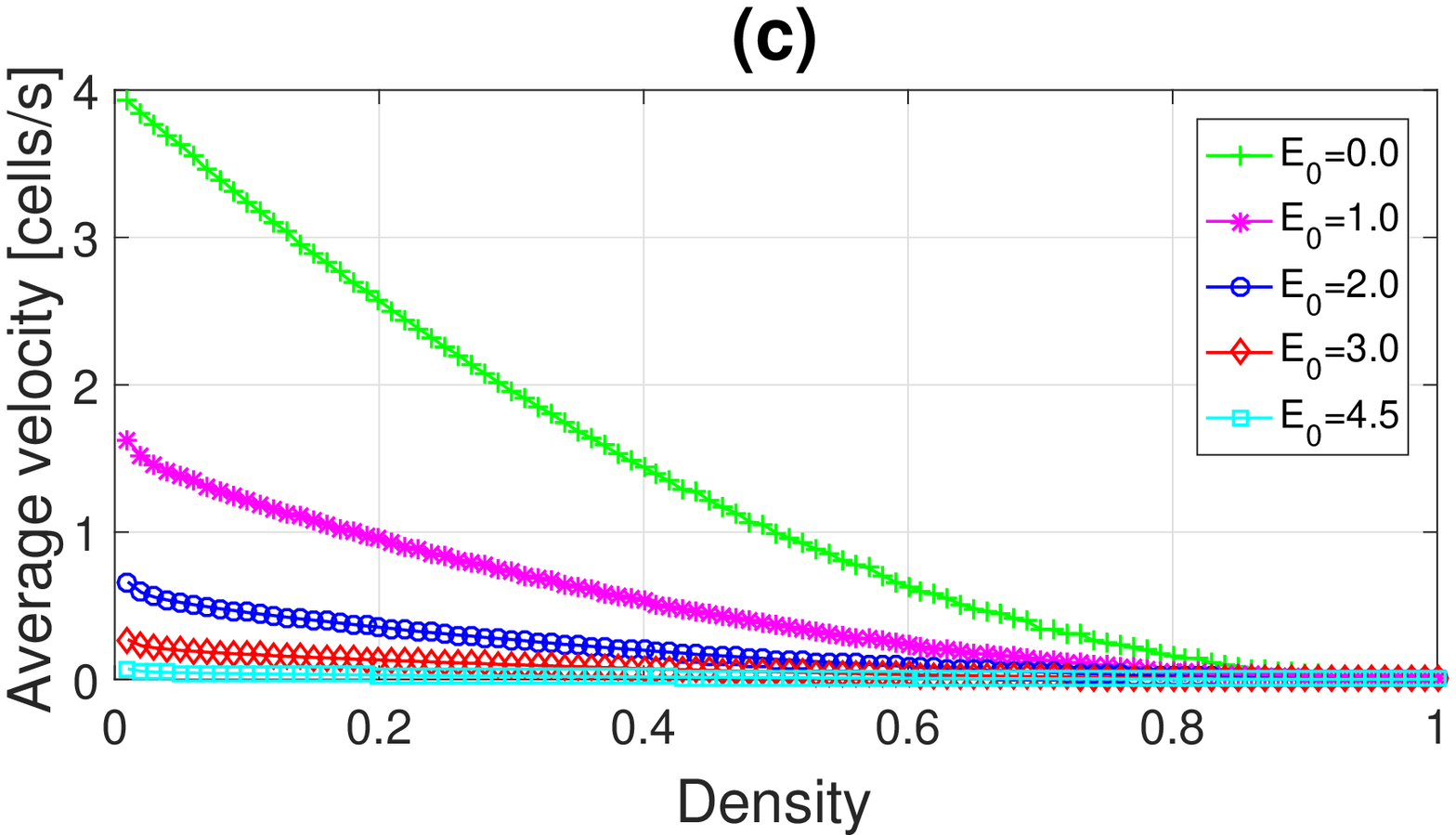} \hfill
\includegraphics[width=.46\textwidth]{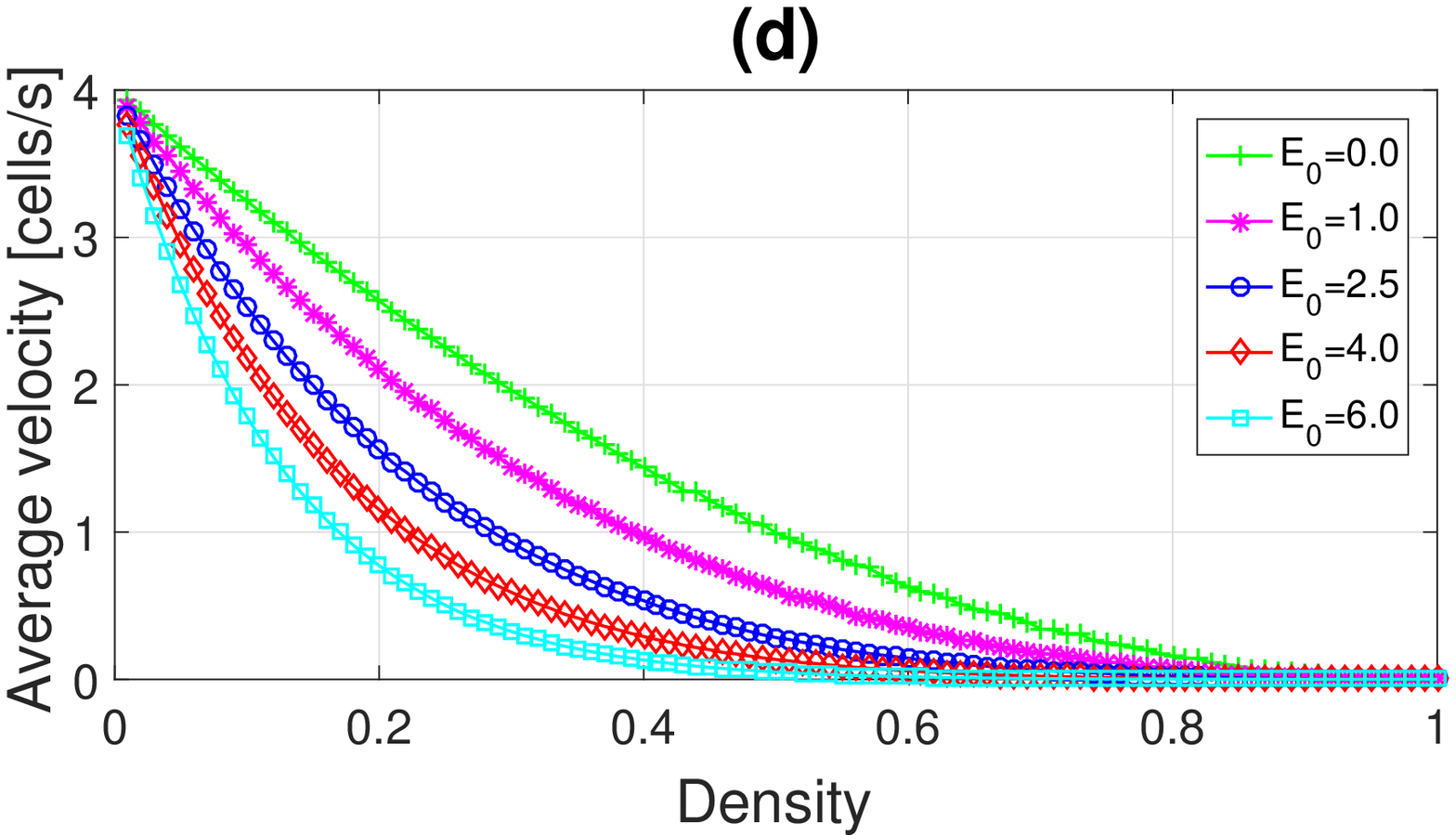}

\vspace{0.2cm}

\hspace{0.2cm} \includegraphics[width=.47\textwidth]{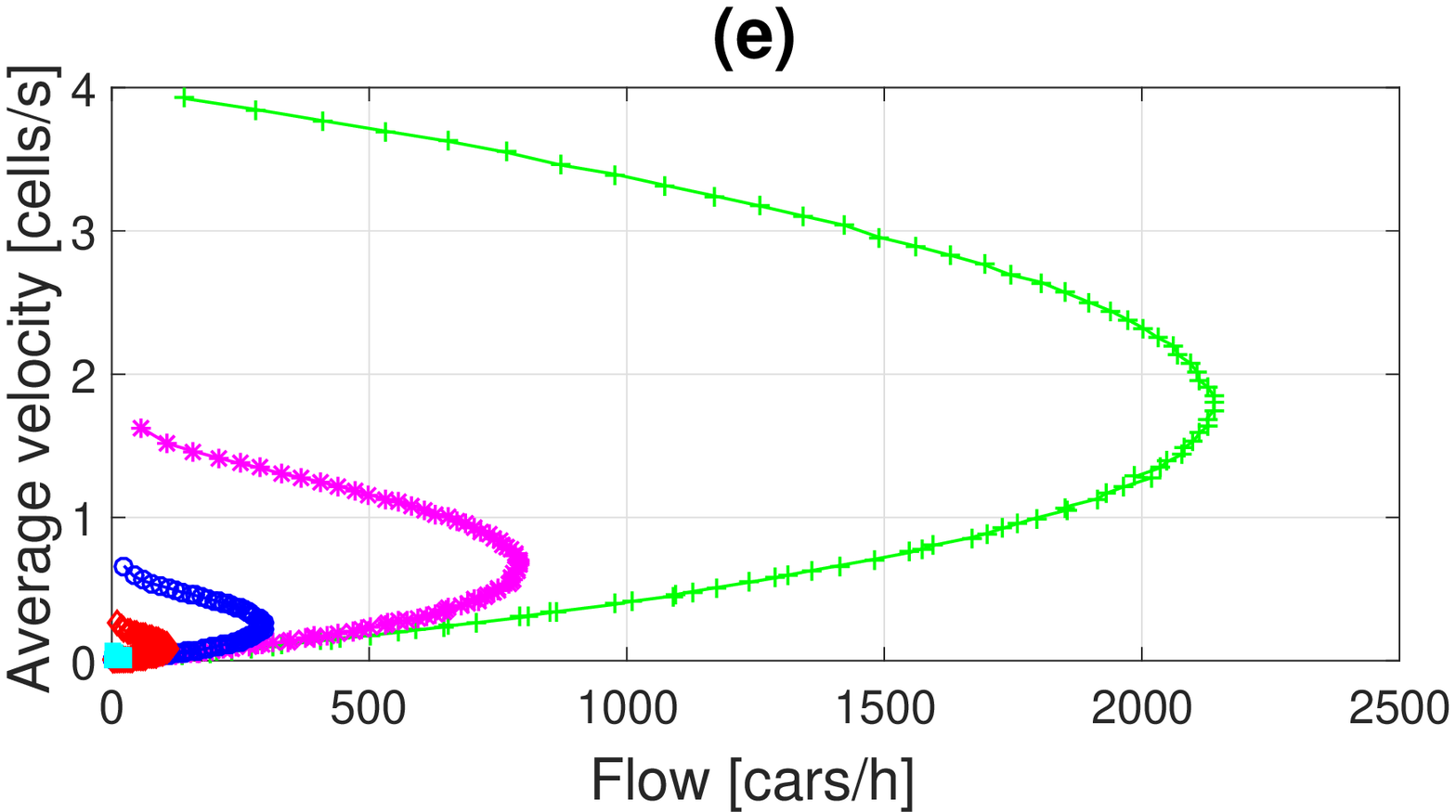} \hfill
\includegraphics[width=.47\textwidth]{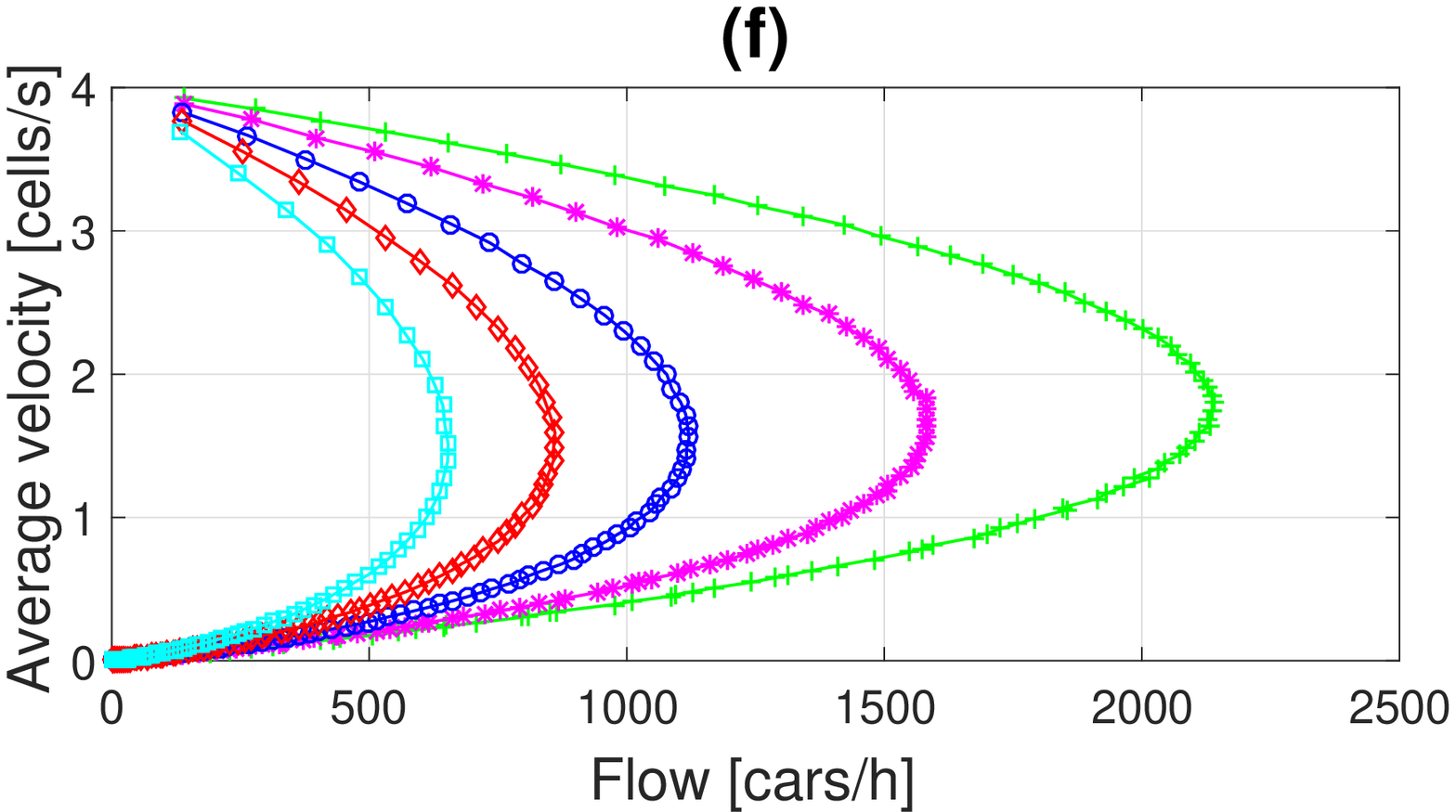}

\caption{Comparison results of the traffic flow on the one-lane highway with five different values of the interaction strength $\Eo$. In all KMC simulations, we take the highway distance of $\approx4.17$ miles ($\approx 6704$m, $\NL=1000$ cells), the look-ahead parameter of $\L=1000$, the multiple move parameter of $\Jump=2$ and the final time of $1$ hour. (a)(b): Long-time averages of the density-flow relationship; (c)(d): Ensemble-averaged velocity of cars versus the density $\bar{\rho}$; (e)(f): Long-time averages of the flow-velocity relationship. (left panel): Results of the first look-ahead rule \eqref{eq:Eb1} with $\Eo=0$ to $4.5$. (right panel): Results of the second look-ahead rule \eqref{eq:Eb2} with $\Eo=0$ to $6.0$. Note that the flux of the KMC simulation in (a) and (b) agrees with the macroscopic averaged flux  \eqref{eq:flux1avg} and \eqref{eq:flux2avg} (shown as the dashed black curves), respectively.}
                                                           \label{fig:flowEL1000}
\end{center}
\end{figure}

However, when the look-ahead distance $\L$ is large, the two look-ahead rules produce different results in all diagrams of the density-flow, density-velocity and flow-velocity relationships as shown in Fig.~\ref{fig:flowEL1000} and in the following Sec.~\ref{subsect.lookdis}.

Take the look-ahead distance $\L=1000$, which is equal to the length $\NL$ of the loop highway. Recall the macroscopic flux under the two rules
\[F=\omega_0\rho(1-\rho)^\Jump e^{-\Eo},\quad \text{and}\quad
F=\omega_0\rho(1-\rho)^\Jump e^{-\bar{\rho}\Eo}.\]
As the dynamics will reach the equilibrium state $\rho(x)\equiv\bar\rho$, the longtime averaged flux should satisfy
\begin{numcases}{\Fbar(\bar{\rho})=}
  \omega_0\bar{\rho}(1-\bar{\rho})^\Jump e^{-\Eo}, \quad
  \text{First rule}\label{eq:flux1avg}\\
  \omega_0\bar{\rho}(1-\bar{\rho})^\Jump e^{-\bar{\rho}\Eo},\quad
  \text{Second rule}   \label{eq:flux2avg}
\end{numcases}
The critical density can then be obtained through a first derivative test.
For the first rule,
\begin{equation}\label{eq:flux1crit}
  \rhoc=\frac{1}{1+\Jump},
\end{equation}
which depends on $\Jump$ but is independent of $\Eo$.
For the second rule,
\begin{equation}\label{eq:flux2crit}
  \rhoc=\frac{2}{(\Eo+\Jump+1)+\sqrt{(\Eo+\Jump+1)^2-4\Eo}},
\end{equation}
which depends on both $\Jump$ and $\Eo$.

In Fig.~\ref{fig:flowEL1000}, we fix multiple move parameter $\Jump=2$, and vary the interaction strength $\Eo$.
Figs.~\ref{fig:flowEL1000}(a) and (b) of the density-flow relationship show that as $\Eo$ increases, the maximum value of the flow $\Fbar$ of both look-ahead rules tend to decrease.
For the first rule \eqref{eq:Eb1} with $\Eo=0$ to $4.5$ shown in Fig.~\ref{fig:flowEL1000}(a), the fluxes of the KMC simulations agree with the corresponding PDE fluxes \eqref{eq:flux1avg} (shown as the dashed black curves). Moreover, all the density-flow curves take their maxima at the same critical density $\rhoc=\frac13$, which is consistent with \eqref{eq:flux1crit}. When $\Eo\geq2.0$, the maximum value of the flow $\Fbar$ becomes very low as the PDE flux decreases exponentially with increasing $\Eo$ indicated in \eqref{eq:flux1avg}.
For the second look-ahead rule \eqref{eq:Eb2} with $\Eo=0$ to $6.0$ shown in Fig.~\ref{fig:flowEL1000}(b), the fluxes of the KMC simulations also agree with the corresponding averaged flux of the PDE model \eqref{eq:flux2avg} (shown as the dashed black curves).  But the value of the critical density $\rhoc$ tends to decrease with increasing $\Eo$, as indicated in \eqref{eq:flux2crit}.
Moreover, the second rule \eqref{eq:Eb2} can still produce relatively larger flux at small values of average density $\bar{\rho}$ than the first rule \eqref{eq:Eb1} does.

The fundamental diagrams of the density-velocity relationship in Figs.~\ref{fig:flowEL1000}(c) and (d) also show differences between two look-ahead rules for the large look-ahead distance $\L$. Fig.~\ref{fig:flowEL1000}(c) for the first rule \eqref{eq:Eb1} shows that as $\Eo$ increases, the value of $\vbar$ at $\bar{\rho}=0.01$ decreases drastically, ranging from the full speed $\vbar=4.0$ cells per second ($\approx 26.8$ m/s or $60$ miles/h) of the case $\Eo=0$ (shown as green ``$+$" signs) to almost $\vbar\approx 0$ of the case $\Eo=4.5$ (shown as cyan squares). On the other hand, Fig.~\ref{fig:flowEL1000}(d) for the second rule \eqref{eq:Eb2} shows that all curves of different cases of $\Eo$ decrease from high values of $\vbar=3.74$ to $4.0$ cells per second ($\approx 25.0$ to $26.8$ m/s or $56.1$ to $60$ miles/h) at $\bar{\rho}=0.01$ and eventually decay to zero with the increasing density $\bar{\rho}$.

Figs.~\ref{fig:flowEL1000}(e) and (f) of the flow-velocity relationship again show big differences between two look-ahead rules for the large look-ahead distance $\L$. Fig.~\ref{fig:flowEL1000}(e) for the first rule \eqref{eq:Eb1} shows that both the magnitude of the flow $\Fbar$ and the range of $\vbar$ decrease with increasing $\Eo$. For the second rule \eqref{eq:Eb2} shown in Fig.~\ref{fig:flowEL1000}(f), only the magnitude of the flow $\Fbar$ decreases as $\Eo$ increases, but both the range of $\vbar$ and the critical value $\vbar_{\rm crit}$ where the flow $\Fbar$ reaches its maximum do not change too much. The value of $\vbar_{\rm crit}$ is around $1.5$ cells per second ($\approx 10$ m/s or $22.5$ miles/h).

\subsection{Numerical comparisons for different look-ahead distances}
\label{subsect.lookdis}

\begin{figure}[!ht]
\begin{center}
\includegraphics[width=.48\textwidth]{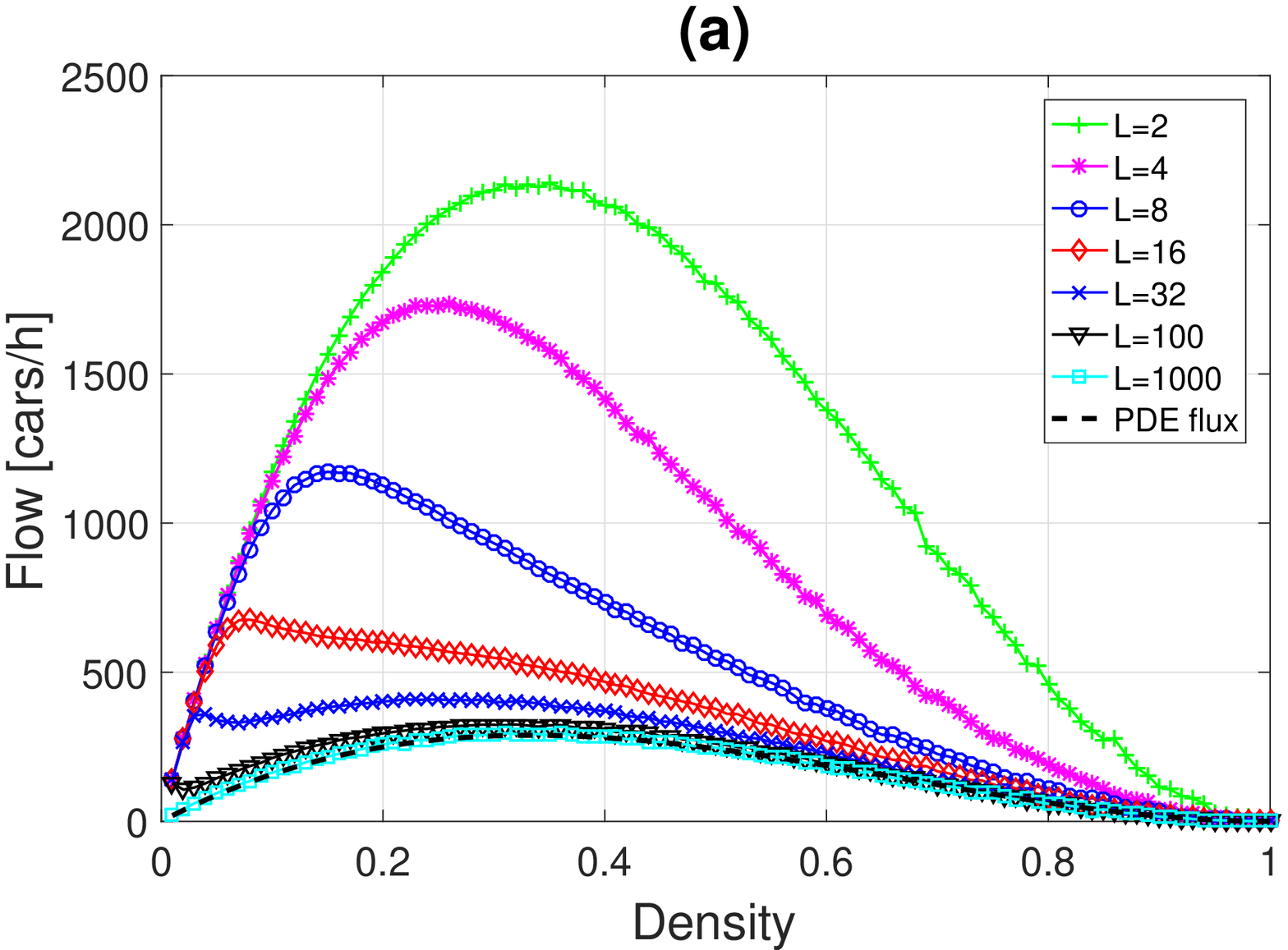} \hfill
\includegraphics[width=.48\textwidth]{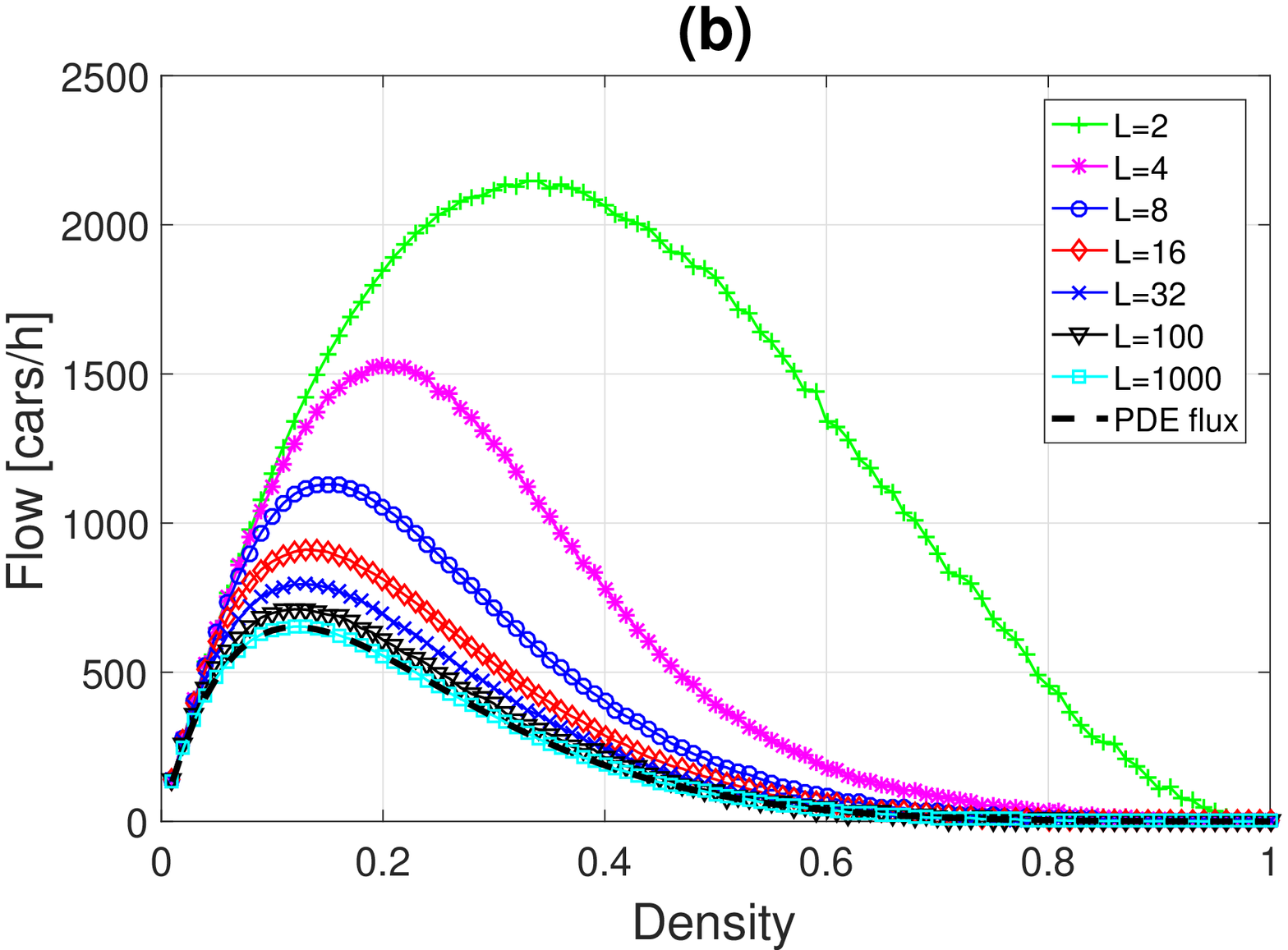}

\vspace{0.2cm}

\hspace{0.2cm} \includegraphics[width=.46\textwidth]{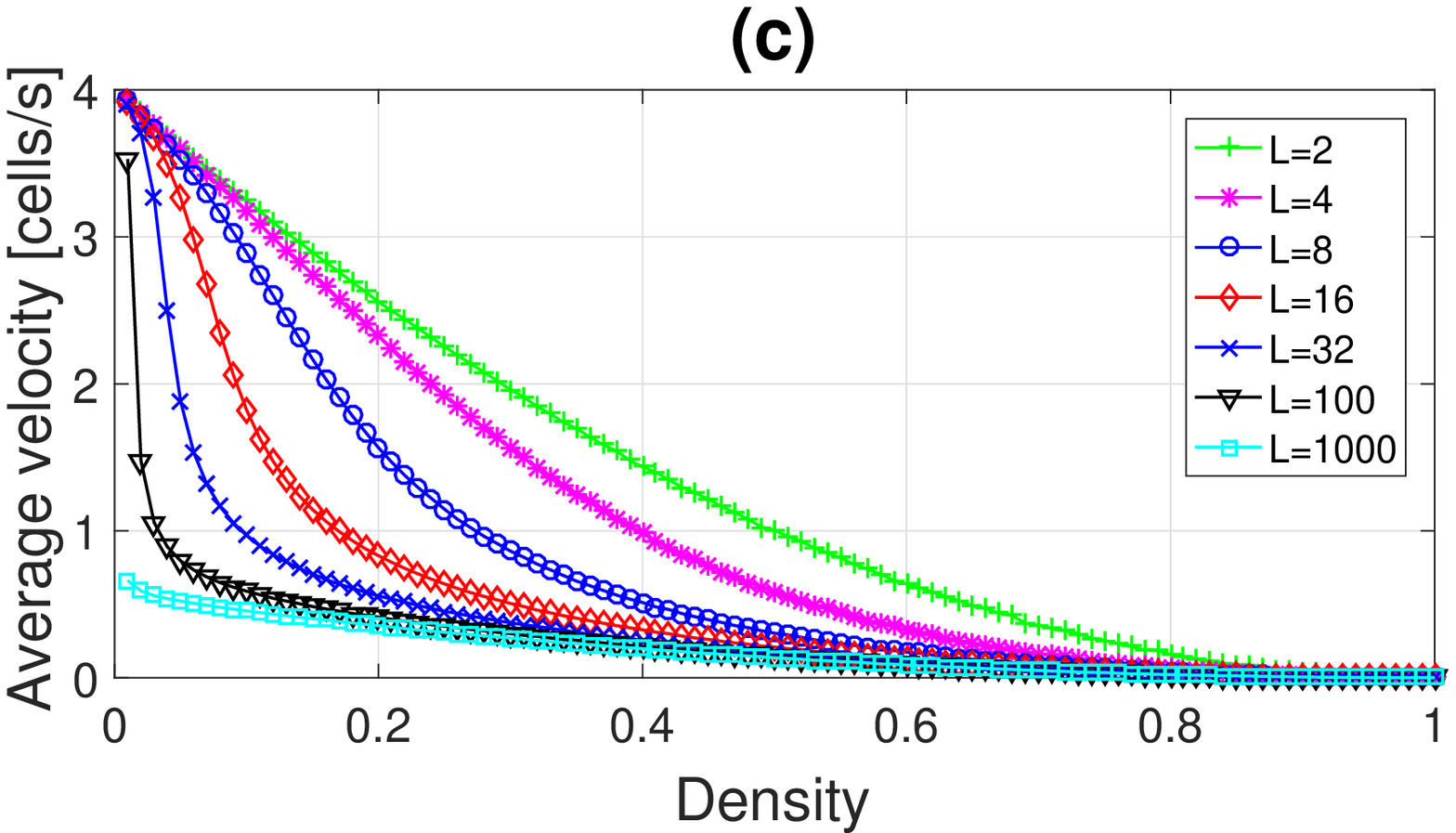} \hfill
\includegraphics[width=.46\textwidth]{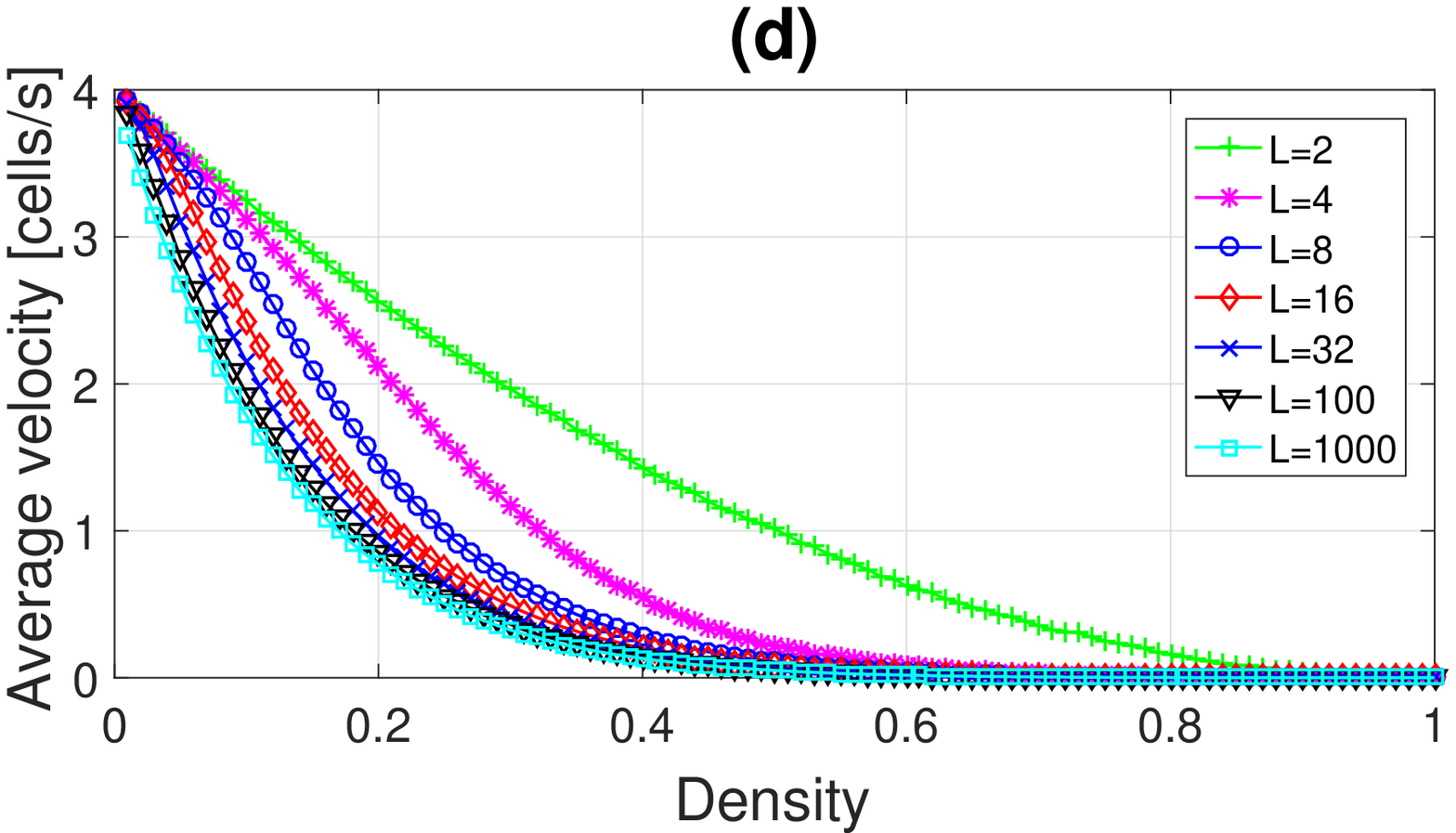}

\vspace{0.2cm}

\hspace{0.2cm} \includegraphics[width=.47\textwidth]{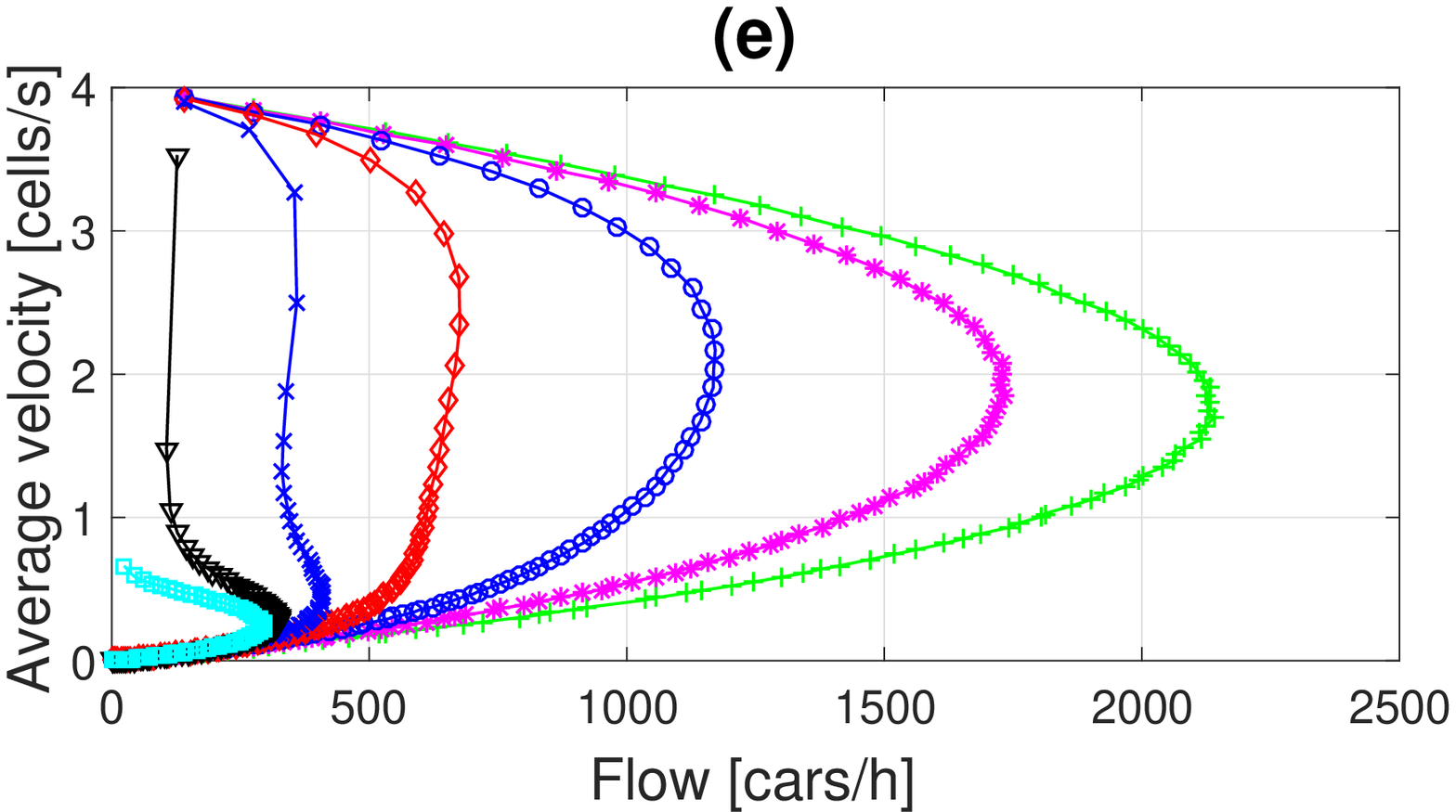} \hfill
\includegraphics[width=.47\textwidth]{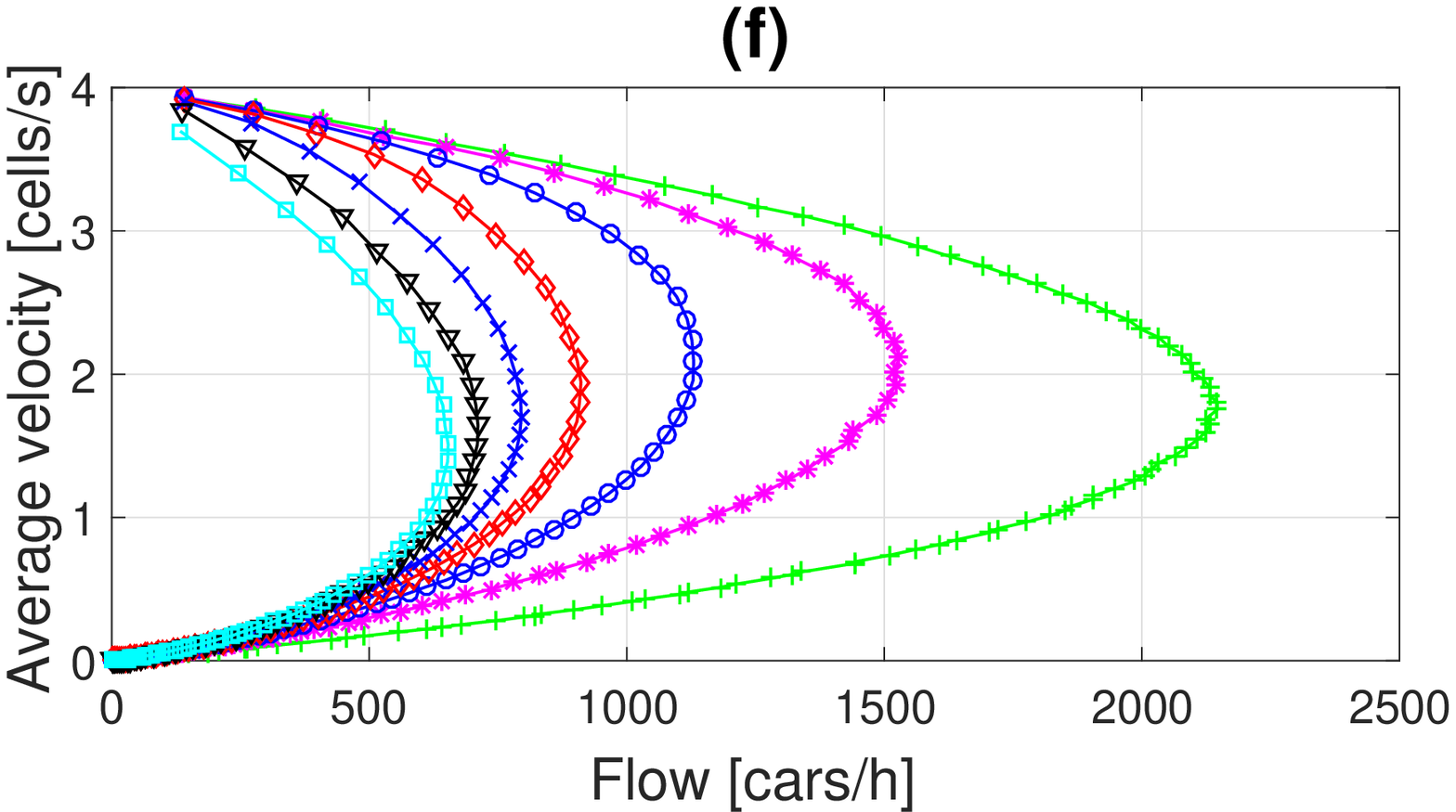}

\caption{Comparison results of the traffic flow on the one-lane highway with seven different values of the look-ahead distance $\L$. In all KMC simulations, we take the highway distance of $\approx4.17$ miles ($\approx 6704$m, $\NL=1000$ cells), the multiple move parameter of $\Jump=2$ and the final time $1$ hour. (a)(b): Long-time averages of the density-flow relationship; (c)(d): Ensemble-averaged velocity of cars versus the density $\bar{\rho}$; (e)(f): Longtime averages of the flow-velocity relationship. (left panel): Results of the first look-ahead rule \eqref{eq:Eb1} with $\Eo=2.0$. (right panel): Results of the second look-ahead rule \eqref{eq:Eb2} with $\Eo=6.0$. Note that for long range interactions ($\L=1000$), the flux of the KMC simulation (shown as cyan squares in (a) and (b)) agrees with the macroscopic averaged flux  \eqref{eq:flux1avg} and \eqref{eq:flux2avg} (shown as the dashed black curve), respectively.}
                                                           \label{fig:flowL}
\end{center}
\end{figure}

Next, we show the effects of the look-ahead distances $\L$ on the flows in more detail in Fig.~\ref{fig:flowL}. For these results, we again take a random car distribution at the initial time on a loop highway of $\approx4.17$ miles ($\approx 6704$m, $\NL=1000$ cells) and observe the behavior of traffic flows as the averaged car density $\bar{\rho}$ increases incrementally from $\bar{\rho}=0.01$ to $\bar{\rho}=0.99$. Here, we use $\Eo=2.0$ for the first look-ahead rule \eqref{eq:Eb1} and $\Eo=6.0$ for the second look-ahead rule \eqref{eq:Eb2}, respectively, and the multiple move parameter $\Jump=2$ for both rules.  All curves exhibit phase transitions between the free-flow phase and the jammed phase. However, Fig.~\ref{fig:flowL} shows that the two look-ahead rules produce different results in all diagrams of the density-flow, density-velocity and flow-velocity relationships when the look-ahead distance $\L$ is large.

Fig.~\ref{fig:flowL}(a) of the density-flow relationship for the first rule \eqref{eq:Eb1} shows that as $\L$ increases, the maximum values of the flow $\Fbar$ tend to decrease, but the value of the critical density $\rhoc$ first decreases ($\L=2$, $4$, $8$ and $16$) and later changes to increase ($\L\geq32$). When $\L=1000$, the flux of the KMC simulations (shown as cyan squares) agrees with the averaged flux for the macroscopic dynamics \eqref{eq:flux1avg} (shown as the dashed black curve). We note that the same KMC simulation results have also been shown as blue circles for $\Eo=2.0$ in Fig.~\ref{fig:flowEL1000}(a). The density-flow curve takes its maximum at the critical density $\rhoc=\frac13$. Following \eqref{eq:flux1avg}, the maximum flux at $\rhoc=\frac13$ is about $3600 \cdot\frac{16}{27} e^{-2.0}\approx289$ cars per hour (recall that $\omega_0=4$s$^{-1}$). Moreover, for a fixed $\bar{\rho}$, the magnitude of the flow $\Fbar$ decreases with increasing $\L$ since the larger is the look-ahead distance the longer is the effective range of interaction between the cars. For the second look-ahead rule \eqref{eq:Eb2} shown in Fig.~\ref{fig:flowL}(b), as $\L$ increases, both the maximum value of the flow $\Fbar$ and the value of the critical density $\rhoc$ tend to decrease.  When $\L=1000$, the flux of the KMC simulations (shown as cyan squares) also matches with the averaged flux of the PDE model \eqref{eq:flux2avg} (shown as the dashed black curve). The same case has also been shown as cyan squares for $\Eo=6.0$ in Fig.~\ref{fig:flowEL1000}(b). We note that even the results of $\L=100$ for both rules (shown as black ``$\triangledown$" signs in Figs.~\ref{fig:flowL}(a) and (b)) are close to the corresponding macroscopic fluxes.

In Figs.~\ref{fig:flowL}(c) and (d), the fundamental diagrams of the density-velocity relationship also show differences between the two look-ahead rules. Fig.~\ref{fig:flowL}(c) for the first rule \eqref{eq:Eb1} shows that as $L$ increases, the ensemble-averaged velocity $\vbar$ decreases very rapidly in the low-density regime and eventually decays to zero with the increasing density $\bar{\rho}$. In particular, while the density-velocity curve of $\L=100$ (shown as black ``$\triangledown$" signs) drops down from a high value of $\vbar=3.53$ cells per second ($\approx 23.7$ m/s or $52.9$ miles/h) at $\bar{\rho}=0.01$, the case of $\L=1000$ (shown as cyan squares) starts from a low value of $\vbar=0.66$ cells per second  ($\approx 4.4$ m/s or $9.9$ miles/h) at $\bar{\rho}=0.01$. We recall that the loop highway has a length of $\NL=1000$ cells, so $\bar{\rho}=0.01$ means that there are a total of only $10$ cars on this highway of $\approx4.17$ miles ($\approx 6704$m). Their average velocity $\langle v \rangle$ is so low, which indicates that the first look-ahead rule \eqref{eq:Eb1} is not reasonable for large look-ahead distances $\L$. On the other hand, Fig.~\ref{fig:flowL}(d) shows that the second look-ahead rule \eqref{eq:Eb2} produces more reasonable results, even for large look-ahead distances $\L$. The case of $\L=1000$ (shown as cyan squares) gradually decreases from a high value of $\vbar=3.70$ cells per second ($\approx 24.7$ m/s or $55.5$ miles/h) at $\bar{\rho}=0.01$.

Figs.~\ref{fig:flowL}(e) and (f) show that when the look-ahead distance $\L=2$, the flow-velocity curves for both rules (shown as green ``$+$" signs) reach their maxima at the critical
velocity $\vbar_{\rm crit} \approx 1.7$ cells per second ($\approx 11.6$ m/s or $25.5$ miles/h). As $\L$ increases in Fig.~\ref{fig:flowL}(e) of the first look-ahead rule \eqref{eq:Eb1}, the maximum value of the flow decreases and the critical value
$\vbar_{\rm crit}$ increases to be higher than $2$ cells per second ($\L=4$, $8$ and $16$).
When $\L=32$, the result (shown as blue ``x" signs) produces two local maxima in the flow. As $\L$ further increases, the range of the average velocity eventually becomes very small ($\L=1000$, shown as cyan squares). On the other hand, Fig.~\ref{fig:flowL}(f) shows that for all $\L$, the flow-velocity curve of the second look-ahead rule \eqref{eq:Eb2} has the full range from a high speed down to zero. As $\L$ increases from $2$, the maximum value of the flow decreases, but the critical value $\vbar_{\rm crit}$ first increases and becomes higher than $2$ cells per second ($\L=2, 4$ and $8$), then decreases to be lower than $2$ cells per second ($\L\geq16$).

\subsection{Numerical comparisons for different multiple move parameters}
\label{subsect.jump}

\begin{figure}[!ht]
\begin{center}
\includegraphics[width=.48\textwidth]{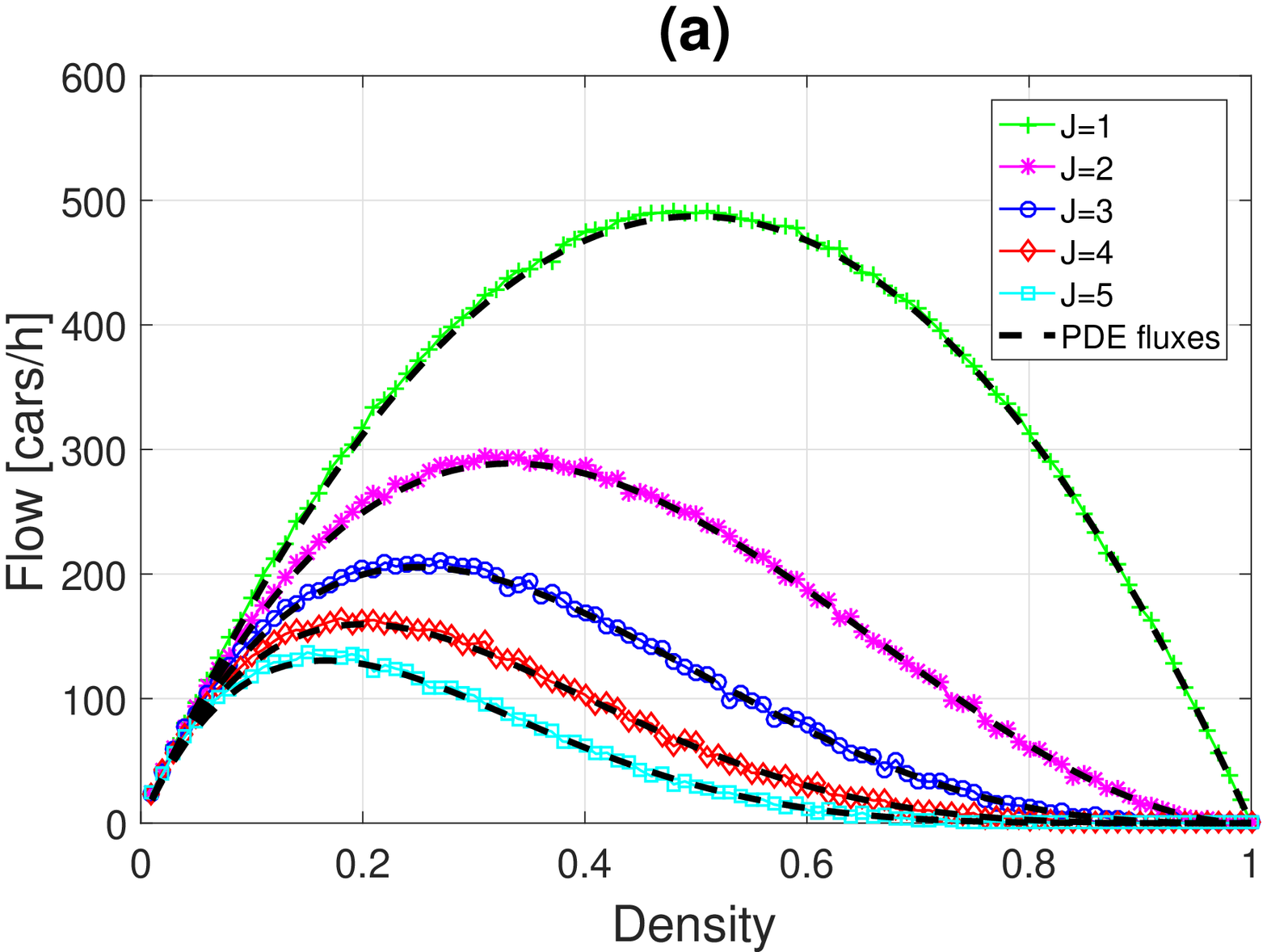} \hfill
\includegraphics[width=.48\textwidth]{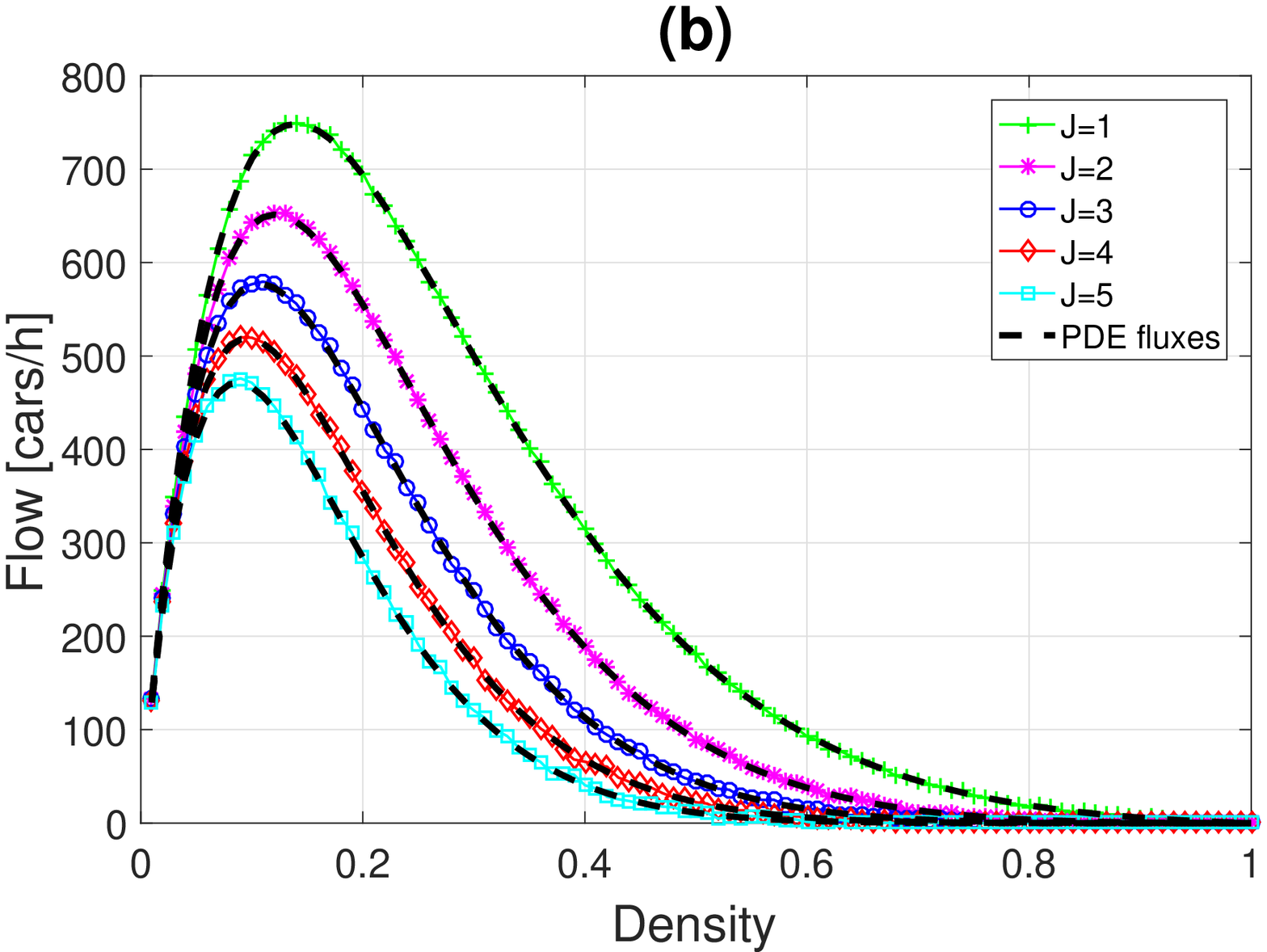}

\vspace{0.2cm}

\hspace{0.2cm} \includegraphics[width=.46\textwidth]{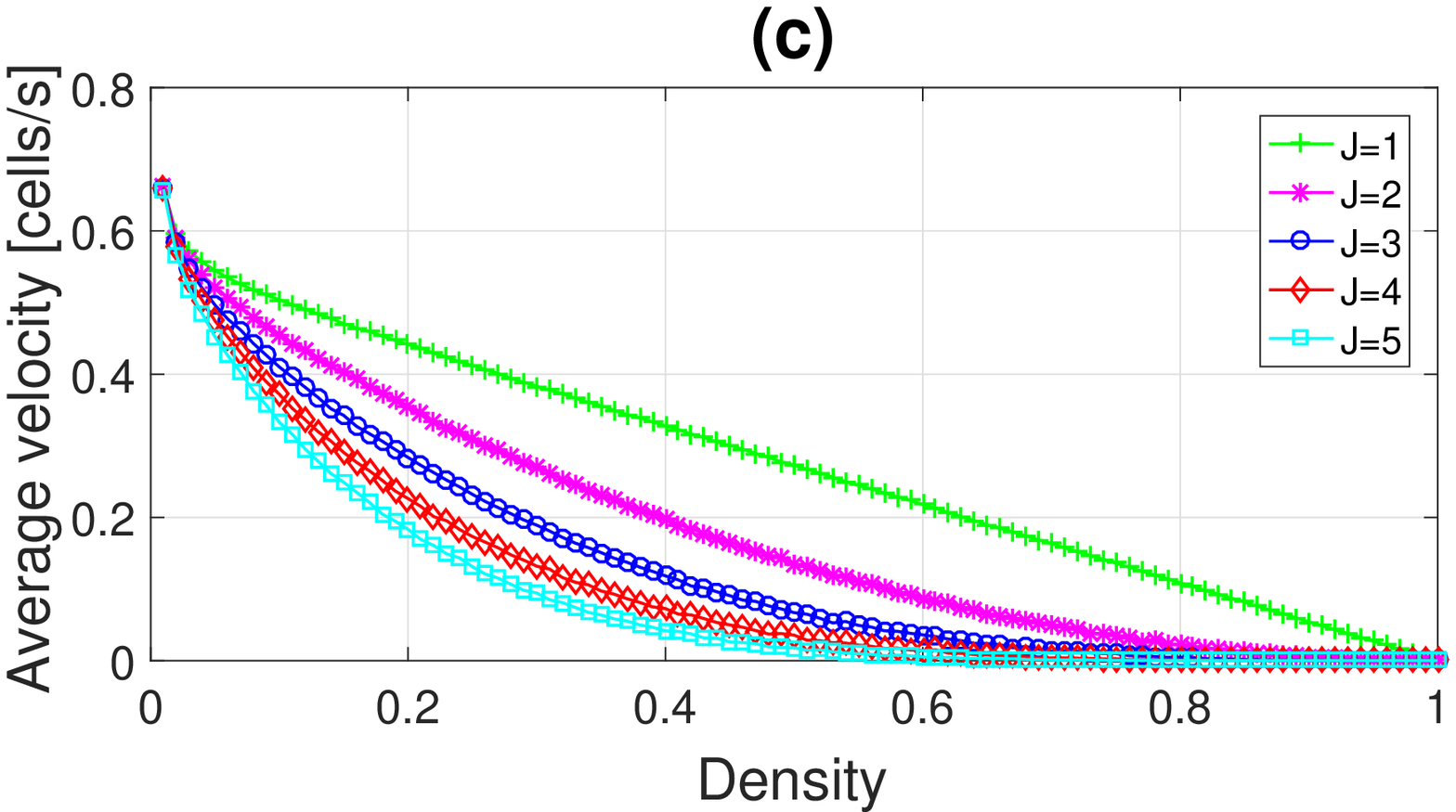} \hfill
\includegraphics[width=.46\textwidth]{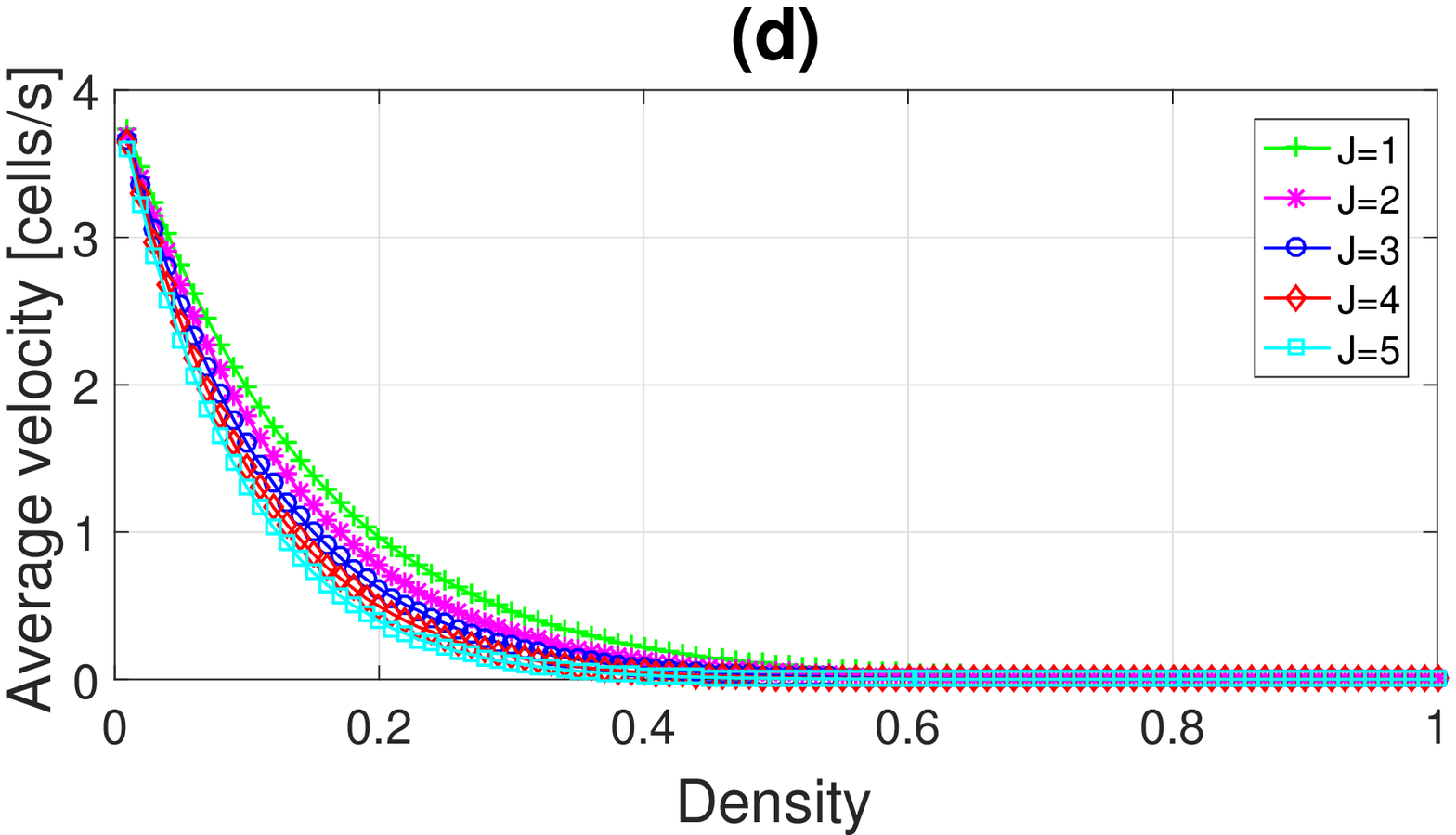}

\vspace{0.2cm}

\hspace{0.2cm} \includegraphics[width=.47\textwidth]{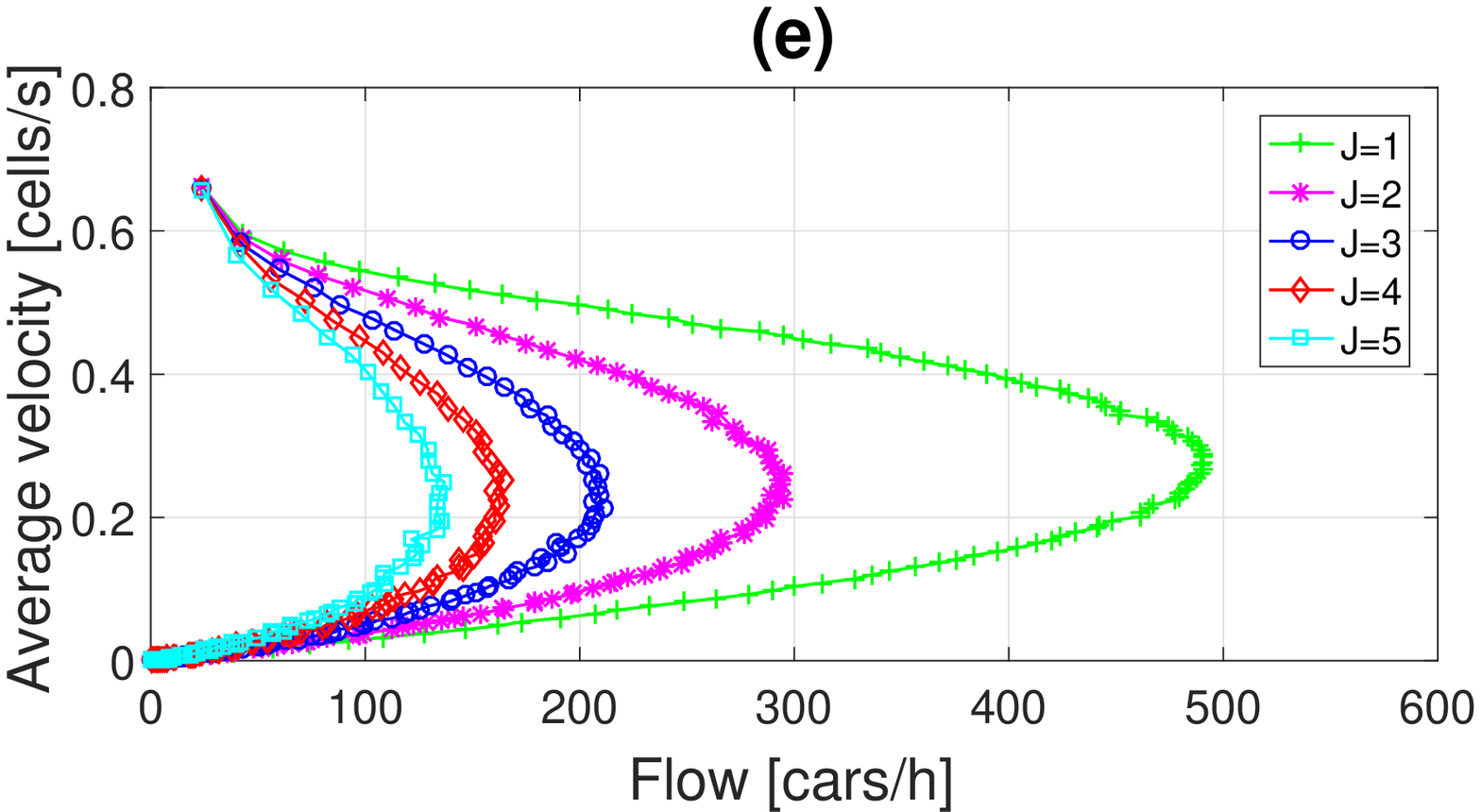} \hfill
\includegraphics[width=.47\textwidth]{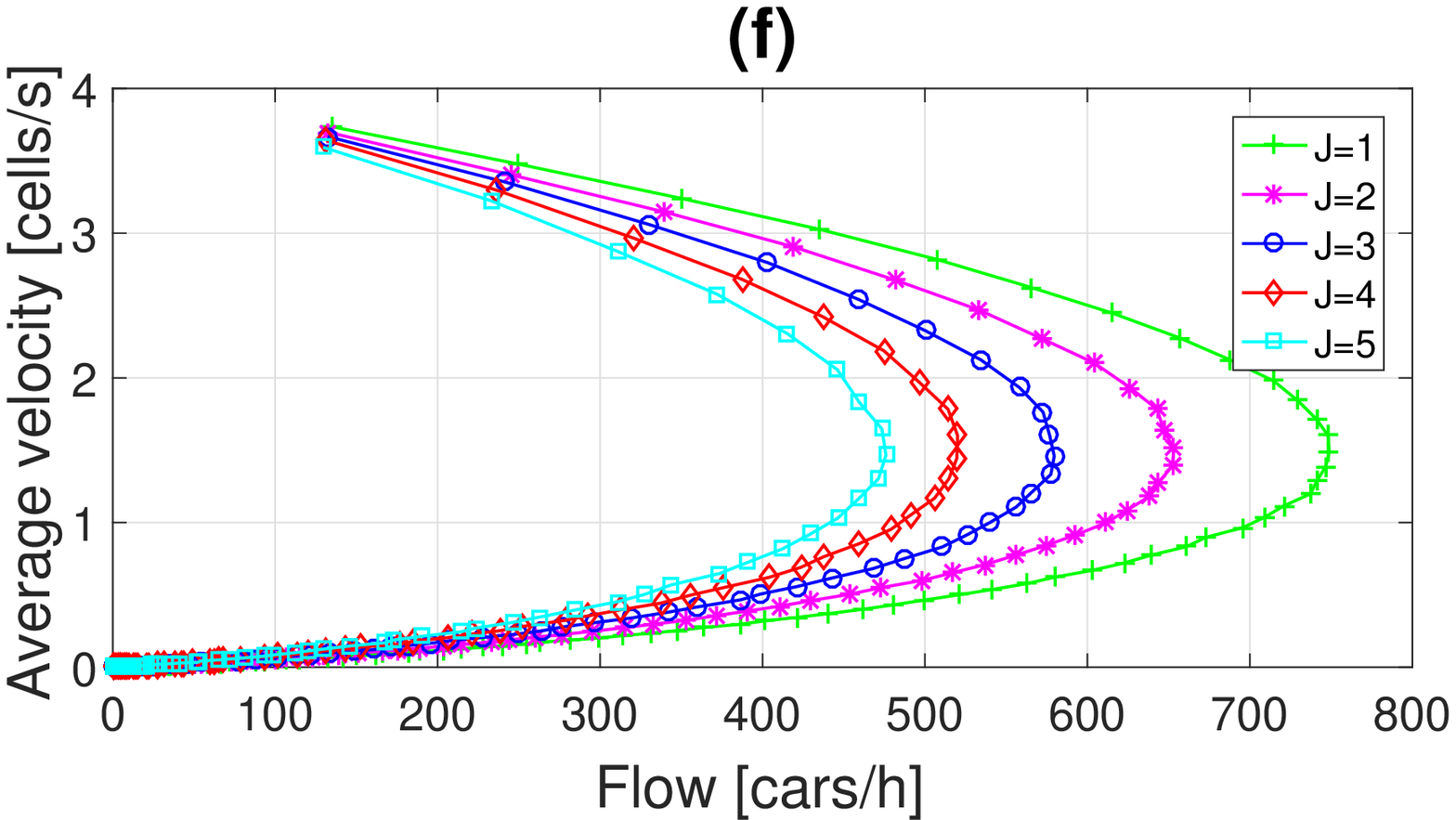}

\caption{Comparison results of the traffic flow on the one-lane highway with five different values of the multiple move parameter $\Jump$. In all KMC simulations, we take the highway distance of $\approx4.17$ miles ($\approx 6704$m, $\NL=1000$ cells), the look-ahead distance of $\L=1000$ and the final time $1$ hour. (a)(b): Long-time averages of the density-flow relationship; (c)(d): Ensemble-averaged velocity of cars versus the density $\bar{\rho}$; (e)(f): Long-time averages of the flow-velocity relationship. (left panel): Results of the first look-ahead rule \eqref{eq:Eb1} with $\Eo=2.0$. (right panel): Results of the second look-ahead rule \eqref{eq:Eb2} with $\Eo=6.0$. Note that for each value of $\Jump=1,2,\ldots, 5$, the flux of the KMC simulation in (a) and (b) agrees with  the macroscopic averaged flux in \eqref{eq:flux1avg} and \eqref{eq:flux2avg} (shown as the dashed black curves), respectively. Also, note the differences in the ranges of $Y$-axis between the left and right panels.}
                                                           \label{fig:flowJ}
\end{center}
\end{figure}

Finally, we show the effects of the multiple move parameter $\Jump$ on the flows in Fig.~\ref{fig:flowJ}. For these results, we still take a random car distribution at the initial time on a loop highway of $\approx4.17$ miles ($\approx 6704$m, $\NL=1000$ cells) and observe the behavior of traffic flows as the averaged car density $\bar{\rho}$ increases incrementally from $\bar{\rho}=0.01$ to $\bar{\rho}=0.99$. Here, we also use $\Eo=2.0$ for the first look-ahead rule \eqref{eq:Eb1} and $\Eo=6.0$ for the second look-ahead rule \eqref{eq:Eb2}, respectively, and take the look-ahead distance of $\L=1000$ for both rules. Fig.~\ref{fig:flowJ} shows the differences between two look-ahead rules with the large look-ahead distance $\L$.

Fig.~\ref{fig:flowJ}(a) shows the density-flow relationship for the first rule \eqref{eq:Eb1} with $\Jump$ increasing from $1$ to $5$. The fluxes match beautifully with the macroscopic averaged flux \eqref{eq:flux1avg} (shown as the dashed black curves). The case $\Jump=1$ corresponds to the LWR type model, where the curve is symmetric and concave. For $\Jump\geq2$, the curves become neither convex nor concave, and have a right-skewed asymmetry. This indicates that our new model is more realistic. Moreover, for a fixed $\bar{\rho}$, the magnitude of the flow $\Fbar$  decreases with increasing $\Jump$, which verifies the heuristic argument in Sec.\ref{sec:micro}. For the second look-ahead rule \eqref{eq:Eb2} shown in Fig.~\ref{fig:flowJ}(b), the microscopic fluxes agree with the macroscopic averaged flux \eqref{eq:flux2avg} very well. The critical density $\rhoc$ is located at \eqref{eq:flux2crit} and it gets smaller as $\Jump$ increases. For instance, when $\Jump=1$ and $\Eo=6.0$ (shown as green ``+" signs), we have $\rhoc=\frac{1}{4+\sqrt{10}}\approx0.14$ and the maximum flux is about $748$ cars per hour.

The fundamental diagrams of the density-velocity relationship in Figs.~\ref{fig:flowJ}(c) and (d) also show differences between two look-ahead rules (note that the ranges of $Y$-axis are different in two figures). Fig.~\ref{fig:flowJ}(c) for the first rule \eqref{eq:Eb1} shows that at the same $\bar{\rho}$, the ensemble-averaged velocity $\vbar$ decreases as $\Jump$ increases. All cases of different $\Jump$ start from a low value of $\vbar=0.66$ cells per second ($\approx 4.4$ m/s or $9.9$ miles/h) at $\bar{\rho}=0.01$. As we pointed out in Fig.~\ref{fig:flowL}(c) in Sec.\ref{subsect.lookdis}, the low average velocity $\langle v \rangle$ at low densities indicates that the first look-ahead rule \eqref{eq:Eb1} is not reasonable for large look-ahead distances $\L$. Here again, Fig.~\ref{fig:flowJ}(d) shows that for large look-ahead distances $\L$, the second look-ahead rule \eqref{eq:Eb2} produces more reasonable results. All curves of different cases of $\Jump$ gradually decrease from a high value of $\vbar=3.74$ cells per second ($\approx 25.0$ m/s or $56.1$ miles/h) at $\bar{\rho}=0.01$ and eventually decay to zero with the increasing density $\bar{\rho}$.

Figs.~\ref{fig:flowJ}(e) and (f) of the flow-velocity relationship show that for a fixed value of $\vbar$, the magnitude of the flow $\Fbar$ decreases with increasing $\Jump$. However, the value of $\vbar_{\rm crit}$ where the flow $\Fbar$ reaches its maximum does not change too much as $\Jump$ increases. For the first look-ahead rule \eqref{eq:Eb1} shown in Fig.~\ref{fig:flowJ}(e), the value of $\vbar_{\rm crit}$ is around $0.25$ cells per second ($\approx 1.67$ m/s or $3.75$ miles/h). For the second rule \eqref{eq:Eb2} shown in Fig.~\ref{fig:flowJ}(f), the value of $\vbar_{\rm crit}$ is around $1.5$ cells per second ($\approx 10$ m/s or $22.5$ miles/h).

\section{Conclusion}\label{sec:conclusion}

We have presented a new class of one-dimensional (1D) models to study traffic flows. Our work is motivated by the growing need to understand mechanisms leading to traffic jams and develop a quantitative approach to the optimal design of transportation systems. The cellular automata (CA) traffic models proposed here incorporate stochastic dynamics for the movement of cars by using the Arrhenius type look-ahead rules of each car, which take into account of the nonlocal slow-down effect. In particular, we considered two different look-ahead rules: the first one is based on the distance from the car under consideration to the car in front of it; the second one depends on the car density ahead. Both rules feature a novel idea of multiple moves, which plays a key role in recovering the right-skewed non-concave flux in the macroscopic dynamics. Through a semi-discrete mesoscopic stochastic process, we derive the coarse-grained macroscopic dynamics of the CA model.

To simulate the proposed CA models, we applied an efficient list-based KMC algorithm with a fast search that can further improve computational efficiency. In the KMC method, the dynamics of cars is described in terms of the transition rates corresponding to possible configurational changes of the system, and then the corresponding time evolution of the system can be expressed in terms of these rates. While the Metropolis Monte Carlo (MMC) method is a way of simulating an equilibrium distribution for a model, the KMC is more suitable for simulating the time evolution of the traffic systems. Moreover, since the KMC algorithm is ``rejection-free'', we choose the KMC as one of our contributions in terms of computational efficiency. The KMC simulations relied on the calibration of model parameters: the characteristic time $\tau_0$, the interaction strength $E_0$, the look-ahead parameter $\L$ and the multiple move parameter $\Jump$. Then we used the KMC simulations to quantitatively predict the time evolution of the traffic flows.

Our numerical results show that the fluxes of the KMC simulations agree with the coarse-grained macroscopic averaged fluxes under various parameter settings. We obtained fundamental diagrams that display several important observed traffic states. In particular, our models capture the right-skewed non-concave asymmetry in the fundamental diagram of the density-flow relationship, which is well-known in realistic traffic measurements \cite{KuG11}. Comparison of the numerical results of the two look-ahead rules shows that in long-range interactions limit with large look-ahead parameter $\L$, the two rules produce different coarse-grained macroscopic averaged fluxes. But for small $\L$, both rules produce similar results.

Physically we do not expect that human drivers would (or even could) have a perception of traffic up front for many cars. Therefore, the look-ahead horizon is typically in the range of $50$ to $150$m, which corresponds to the small-to-intermediate values of $\L=8$ to $24$ and both rules exhibit reasonable behavior in this regime. However, with the fast development of self-driving vehicles equipped with vehicle-to-vehicle communication and a variety of techniques to perceive their surroundings, such as radar, Lidar, sonar, odometry and GPS \cite{Thr10}, we/cars may ``look" far ahead to reach large $\L$. In that situation, the second look-ahead rule may be more suitable than the first one.

As one of our main goals is to compare the two look-ahead rules, we propose our CA models in a closed system and take the periodic boundary conditions to keep the number of cars and the density constant in a single simulation. Therefore, we have not applied our models to simulate some more complex non-stationary features, such as traffic breakdowns at bottlenecks \cite{Ker04}. It is possible to improve the models further in the following directions. We can include entrances and exits in the models by adding dynamical mechanisms such as adsorption/desorption. In reality, there are multi-lanes on highways and fast vehicles may change lanes to bypass slow ones. We also need to consider different types of vehicles, such as cars and trucks with unequal sizes and speeds. More complicated models addressing these aspects will be explored in the future.

\section*{Acknowledgment}
YS is partially supported by the NSF Grants DMS-1620212, DMS-1913146 and a SC EPSCoR GEAR Award.  CT is partially supported by the NSF Grant DMS-1853001.



\vskip 12 pt

\begin{thebibliography}{10}

\bibitem{AlS08}
T.~Alperovich and A.~Sopasakis,
Stochastic description of traffic flow,
{\it J. Stat. Phys.} {\bf 133} (2008) 1083--1105.

\bibitem{BHH95}
M.~Bando, K.~Hasebe, A.~Nakayama, A.~Shibata and Y.~Sugiyama,
Dynamical model of traffic congestion and numerical simulation, {\it Phys. Rev. E} {\bf 51} (1995) 1035--1042.

\bibitem{BSS98}
R.~Barlovic, L.~Santen, A.~Schadschneider and M.~Schreckenberg, Metastable states in cellular automata for traffic flow,
{\it Eur. J. Phys. B} {\bf 5} (1998) 793--800.

\bibitem{BeD11}
N.~Bellomo and C.~Dogbe,
On the modeling of traffic and crowds: A survey of models, speculations, and perspectives,
{\it SIAM Rev.} {\bf 53} (2011) 409--463.

\bibitem{BML92}
O.~Biham, A.~A.~Middleton and D.~Levine,
Self-organization and dynamical transition in traffic-flow models,
{\it Phys. Rev. A} {\bf 46} (1992) R6124-6127.

\bibitem{BBS95}
J.~L.~Blue, I.~Beichl and F.~Sullivan,
Faster Monte Carlo simulations,
{\it Phys. Rev. E} {\bf 51} (1995) R867.

\bibitem{BKL75}
A.~B.~Bortz, M.~H.~Kalos and J.~L.~Lebowitz,
A new algorithm for Monte Carlo simulation of Ising spin systems,
{\it J. Comput. Phys.} {\bf 17} (1975) 10--18.

\bibitem{CKPT}
A.~Chertock, A.~Kurganov, A.~Polizzi and I.~Timofeyev,
Pedestrian flow models with slowdown interactions,
{\it Math. Models Methods Appl. Sci.} {\bf 24} (2014) 249--275.

\bibitem{CSS00}
D.~Chowdhury, L.~Santen and A.~Schadschneider,
Statistical physics of vehicular traffic and some related systems,
{\it Phys. Rep.} {\bf 329} (2000) 199--329.

\bibitem{CrL86}
M.~Cremer and J.~Ludwig,
A fast simulation model for traffic flow on the basis of boolean operations,
{\it Math. Comput. Simulation} {\bf 28} (1986) 297--303.

\bibitem{DuS07}
N.~Dundon and A.~Sopasakis,
Stochastic modeling and simulation of multi-lane traffic,
{\it Transp. Traffic Theory} {\bf 17} (2007) 661--691.

\bibitem{GHR61}
D.~C.~Gazis, R.~Herman and R.~ W.~ Rothery,
Nonlinear follow-the-leader models of traffic flow,
{\it Oper. Res.}, {\bf 9} (1961) 545--567.

\bibitem{Gre35}
B.~D.~Greenshields,
A study of traffic capacity,
{\it Proc. Highw. Res. Board} {\bf 14} (1935) 448--477.

\bibitem{HNS03}
K.~Hasebe, A.~Nakayama and Y.~Sugiyama,
Dynamical model of a cooperative driving system for freeway traffic,
{\it Phys. Rev. E} {\bf 68} (2003) 026102.

\bibitem{HNS04}
K.~Hasebe, A.~Nakayama and Y.~Sugiyama,
Equivalence of linear response among extended optimal velocity models,
{\it Phys. Rev. E} {\bf 69} (2004) 017103.

\bibitem{HST14}
C.~Hauck, Y.~Sun and I.~Timofeyev,
On cellular automata models of traffic flow with look ahead potential,
{\it Stoch. Dynam.} {\bf 14} (2014) 1350022.

\bibitem{Hel01}
D.~Helbing,
Traffic and related self-driven many-particle systems,
{\it Rev. Mod. Phys.} {\bf 73} (2001) 1067--1141.

\bibitem{HHS02}
D.~Helbing, A.~Hennecke, V.~Shvetsov and M.~Treiber,
Micro- and macro-simulation of freeway traffic,
{\it Math. Comput. Modell.} {\bf 35} (2002) 517--547.

\bibitem{Ker04}
B.~S.~Kerner, {\it The Physics of Traffic}, (Springer, Heidelberg, 2004).

\bibitem{KeR97}
B.~S.~Kerner and H.~Rehborn,
Experimental properties of phase transitions in traffic flow,
{\it Phys. Rev. Lett.} {\bf 79} (1997) 4030--4033.

\bibitem{KSS00}
W.~Knospe, L.~Santen, A.~Schadschneider and M.~Schreckenberg, Towards a realistic microscopic description of highway traffic,
{\it J. Phys. A} {\bf 33} (2000) L477-L485.

\bibitem{KSS02}
W.~Knospe, L.~Santen, A.~Schadschneider and M.~Schreckenberg,
Single-vehicle data of highway traffic: Microscopic description of traffic phases,
{\it Phys. Rev. E} {\bf 65} (2002) 056133.

\bibitem{KuG11}
R.~K\"{u}hne and N.~H.~Gartner (Eds.),
{\it 75 Years of the Fundamental Diagram for Traffic Flow
Theory: Greenshields Symposium},
(Transportation Research Board E-Circular, 2011).

\bibitem{KP09}
A.~Kurganov and A.~Polizzi,
Non-oscillatory central schemes for traffic flow models with Arrhenius look-ahead dynamics,
{\it Netw. Heterog. Media}, {\bf 4} (2009) 431--451.

\bibitem{Lee18}
Y.~Lee,
Wave breaking in a class of non-local conservation laws,
{\it arXiv:1812.10406}, (2018).

\bibitem{Lee19}
Y.~Lee,
Thresholds for shock formation in traffic flow models with nonlocal-concave-convex flux,
{\it J. Differ. Equ.}, {\bf 266} (2019) 580--599.

\bibitem{LL15}
Y.~Lee and H.~Liu,
Thresholds for shock formation in traffic flow models with Arrhenius look-ahead dynamics,
{\it DCDS-A}, {\bf 35} (2015) 323--339.

\bibitem{LT19}
Y.~Lee and C.~Tan,
A sharp critical threshold for a traffic flow model with look-ahead dynamics,
{\it arXiv:1905.05090}, (2019).

\bibitem{LWS99}
H.~Lenza, C.~K.~Wagner and R.~Sollacher,
Multi-anticipative car-following model,
{\it Eur. Phys. J. B} {\bf 7}  (1999) 331--335.

\bibitem{LiT05}
T.~Li,
Nonlinear dynamics of traffic jams,
{\it Physica D} {\bf 207} (2005) 41--51.

\bibitem{Lig85}
T.~D.~Liggett, {\it Interacting Particle Systems}, (Springer, Berlin, 1985).

\bibitem{LiW55}
M.~Lighthill and G. Whitham,
On kinematic waves II. A theory of traffic flow on long crowded roads,
{\it Proc. Roy. Soc. London, Ser. A.} {\bf 229} (1955) 317--345.

\bibitem{MaD05}
S.~Maerivoet and B.~De Moor,
Cellular automata models of road traffic,
{\it Phys. Rep.} {\bf 419} (2005) 1--64.

\bibitem{May90}
A.~D.~May,
{\it Traffic Flow Fundamentals} (Prentice-Hall, Englewood Cliffs, NJ, 1990).

\bibitem{MRR53}
N.~Metropolis,  A.~E.~Rosenbluth,  M.~N.~Rosenbluth,  A.~H.~Teller and E.~Teller,
Equation of state calculations by fast computing machines,
{\it J. Chem. Phys.} {\bf 21} (1953) 1087--1092.

\bibitem{Nag99}
T.~Nagatani,
Stabilization and enhancement of traffic flow by the next-nearest-neighbor interaction,
{\it Phys. Rev. E} {\bf 60} (1999) 6395--6401.

\bibitem{Nag02}
T.~Nagatani, The physics of traffic jams,
{\it Rep. Prog. Phys.} {\bf 65} (2002) 1331--1386.

\bibitem{NER00}
K.~Nagel, J.~Esser and M.~Rickert, {\it Large-scale traffic simulations for transportation planning}, in Annual Reviews of Computational Physics, edited by D. Stauffer, World Scientific, Singapore, 2000, 151--202.

\bibitem{NaP95}
K.~Nagel and M.~Paczuski,
Emergent traffic jams,
{\it Phys. Rev. E} {\bf 51} (1995) 2909-2918.

\bibitem{NaS92}
K.~Nagel and M.~Schreckenberg,
A cellular automaton model for freeway traffic,
{\it J. Phys. I France} {\bf 2} (1992) 2221--2229.

\bibitem{NSH02}
A.~Nakayama, Y.~Sugiyama and K.~Hasebe,
Effect of looking at the car that follows in an optimal velocity model of traffic flow,
{\it Phys. Rev. E} {\bf 65} (2002) 016112.

\bibitem{Nel06}
P.~Nelson,
On driver anticipation, two-regime flow, fundamental diagrams, and kinematic-wave theory,
{\it Transp. Sci.} {\bf 40} (2006) 165--178.

\bibitem{Scha02}
A.~Schadschneider,
Traffic flow: a statistical physics point of view,
{\it Physica A} {\bf 313} (2002) 153--187.

\bibitem{SCN11}
A.~Schadschneider, D.~Chowdhury and K.~Nishinari,
{\it Stochastic Transport in Complex Systems},
(Elsevier, The Netherlands, 2011).

\bibitem{Sch02}
T.~P.~Schulze,
Kinetic Monte Carlo simulations with minimal searching,
{\it Phys. Rev. E} {\bf 65} (2002) 036704.

\bibitem{SKa06}
A.~Sopasakis and M.~A.~Katsoulakis,
Stochastic modeling and simulation of traffic flow: Asymmetric single exclusion process with Arrhenius look-ahead dynamics,
{\it SIAM J. Appl. Math.} {\bf 66} (2006) 921--944.


\bibitem{SCE11}
Y.~Sun, R.~Caflisch and B.~Engquist,
A multiscale method for epitaxial growth,
{\it SIAM Multiscale Model. Simul.} {\bf 9} (2011) 335--354.

\bibitem{SuT14}
Y.~Sun and I.~Timofeyev,
Kinetic Monte Carlo simulations of one-dimensional and two-dimensional traffic flows: Comparison of two look-ahead rules,
{\it Phys. Rev. E} {\bf89} (2014) 052810.

\bibitem{Thr10}
S.~Thrun,
Toward robotic cars,
{\it Communications of the ACM} {\bf 53} (2010) 99--106.

\bibitem{TrK13}
M.~Treiber and A.~Kesting,
{\it Traffic Flow Dynamics: Data, Models and Simulation},
(Springer, Heidelberg, 2013).

\bibitem{Und61}
R.~T.~Underwood,
{\it Speed, volume, and density relationships: Quality and theory of traffic flow},
(Yale Bureau of Highway Traffic, 1961).

\bibitem{Whi74}
G.~B.~Whitham,
{\it Linear and Nonlinear Waves} (Wiley-Interscience, New York, 1974).

\bibitem{Wie95}
R.~Wiedemann, in {\it Beitr\"{a}ge zur Theorie des Stra{\ss}enverkehrs}, edited by H. Keller (Forschungsgesellschaft f\"{u}r Stra{\ss}en- und Verkehrswesen, K\"{o}ln, 1995).

\bibitem{WBH04}
R.~E.~Wilson, P.~Berg, S.~Hooper and G.~Lunt,
Many-neighbour interaction and non-locality in traffic models, {\it Eur. Phys. J. B} {\bf 39} (2004) 397--408.

\bibitem{Wol86}
S. Wolfram,
{\it Theory and Applications of Cellular Automata},
(World Scientific, Singapore, 1986).

\bibitem{Wol94}
S. Wolfram,
{\it Cellular Automata and Complexity},
(Addison-Wesley, Reading, MA, 1994).


\end{thebibliography}
\end{document}